\documentclass[12pt, draftclsnofoot, onecolumn]{IEEEtran}
\usepackage{yfonts}

\usepackage{fancyhdr}

\usepackage[noadjust]{cite}
\usepackage{epstopdf}
\usepackage{amsthm}
\usepackage{tabularx}
\usepackage{comment}
\usepackage{graphicx}
\usepackage{tikz}
\usepackage[caption=false]{subfig}
\usepackage{amsfonts}
\usepackage{amssymb}
\usepackage{amsmath}
\usepackage{mathtools}
\usepackage{float}
\usepackage{braket}
\usepackage{hyperref}
\hypersetup{
	bookmarksdepth=-2,
	linkcolor=black,
	colorlinks=true,
	citecolor=blue
}
\usepackage[mathscr]{eucal} 

\definecolor{clr_my_dgray}{rgb}{0.4, 0.4, 0.4}
\definecolor{clr_steel_yello}{RGB}{239, 174, 24}
\definecolor{clr_steel_blue}{RGB}{0, 82, 156}
\definecolor{clr_steel_red}{RGB}{199, 3, 45}
\newcommand{\mysquare}[1]{\tikz{\draw[draw=#1] (0,0) rectangle (0.7em,0.7em);}}
\newcommand{\mycircle}[1]{\tikz{\draw[draw=#1] (0,0) circle [radius=0.35em];}}
\newcommand{\mytriangle}[1]{\tikz{\draw[draw=#1] (0,0) -- (0.7em,0) -- (0.35em,0.7em) -- cycle;}}

\makeatletter
\def\@IEEEsectpunct{:\ \,}
\def\paragraph{\@startsection{paragraph}{4}{\z@}{1.5ex plus 1.5ex minus 0.5ex}%
{0ex}{\normalfont\normalsize\bfseries}}
\makeatother

\def\opa/{\text{Operator A}}
\def\opb/{\text{Operator B}}
\def\opc/{\text{Operator C}}
\def\opd/{\text{Operator D}}
\def\cmai/{\text{Atlanta}}
\def\cmaii/{\text{Boston}}
\def\cmaiii/{\text{Phoenix}}

\newcommand{\figref}[1]{Figure~\ref{#1}}
\newcommand{\secref}[1]{Section~\ref{#1}}

\theoremstyle{definition} 

\newtheorem{proposition}{Proposition}
\newtheorem{corollary}{Corollary}
\newcommand\numberthis{\addtocounter{equation}{1}\tag{\theequation}}

\newcommand{\diff}{\mathrm{d}}
\newcommand{\overbar}[1]{\mkern 1.5mu\overline{\mkern-1.5mu#1\mkern-1.5mu}\mkern 1.5mu}
\newcommand{\medcap}{\mathbin{\scalebox{1.5}{\ensuremath{\cap}}}}

\newcommand{\prob}[1]{\text{P}\left(#1\right)}
\newcommand{\indic}[1]{\mathbf{1}_{\scriptstyle #1}}
\newcommand{\expect}[1]{\mathbf{E}\left[#1\right]}
\newcommand{\cov}[2]{\textrm{Cov}\left(#1,\,#2\right)}
\newcommand{\lapof}[2]{\mathcal{L}_{#1}\left(#2\right)}

\newcommand{\sfsinr}{\textsf{SINR}}
\newcommand{\sfrate}{\textsf{Rate}}
\newcommand{\sfgmax}{\textsf{G}}
\newcommand{\sfgmin}{\textsf{g}}

\newcommand{\op}{\mathcal{O}}
\newcommand{\netw}{\textrm{Network 1}}
\newcommand{\nett}{\textrm{Network 2}}

\newcommand{\stra}{\text{FID}}
\newcommand{\strb}{\text{FCD}}

\newcommand{\tauof}[2]{\tau(#1,#2)}

\newcommand{\lamhat}{\hat{\lambda}}
\newcommand{\lamwhat}{\hat{\lambda}_1}
\newcommand{\lamthat}{\hat{\lambda}_2}

\newcommand{\rhohat}{\hat{\rho}}

\newcommand{\pcovw}[1]{\text{P}_\text{c}\left(#1\right)}
\newcommand{\pcovt}[2]{\text{P}_\text{c}\left(#1;\,#2\right)}

\newcommand{\expectw}[2]{\mathbf{E}_{#1}\left[#2\right]}
\newcommand{\Expect}[1]{\mathbf{E}\left[#1\right]}
\newcommand{\expectno}[1]{\mathbf{E}\,#1}
\newcommand{\Exp}{\mathbf{E}}

\newcommand{\lap}[1]{\mathcal{L}_{#1}}
\newcommand{\lapvar}[1]{\mathcal{L}{\left[#1\right]}}
\newcommand{\lapvarof}[2]{\mathcal{L}{\left[#1\right]}\left(#2\right)}

\newcommand{\los}{\mathscr{L}}
\newcommand{\nlos}{\mathscr{N}}
\newcommand{\tlos}{\mathscr{T}}

\newcommand{\assoc}{\mathcal{A}}
\newcommand{\ess}{\mathcal{S}}
\newcommand{\essp}{\mathcal{S'}}
\newcommand{\tee}{\mathcal{T}}
\newcommand{\peeof}[1]{\mathscr{P}(#1)}
\newcommand{\ball}[1]{B_0(#1)}
\newcommand{\ballc}[1]{B_0^c(#1)}
\newcommand{\winofn}{\mathcal{W}^{(n)}}

\newcommand{\expp}[1]{\exp\left(#1\right)}
\newcommand{\cst}{{2\pi\lambda}}
\newcommand{\vol}[1]{\textrm{vol}\left(#1\right)}

\newcommand{\fr}[2]{{\usefont{OMS}{cmr}{m}{n}\selectfont f}_R\left(#1;\,#2\right)}
\newcommand{\fro}[2]{{\usefont{OMS}{cmr}{m}{n}\selectfont f}^o_R\left(#1;\,#2\right)}
\newcommand{\pvoid}[2]{\mu_{#1}\left(#2\right)}

\newcommand{\alos}{\alpha_\text{L}}
\newcommand{\anlos}{\alpha_\text{N}}
\newcommand{\clos}{c_\text{L}}
\newcommand{\cnlos}{c_\text{N}}
\newcommand{\plos}{p_\text{L}}
\newcommand{\pnlos}{p_\text{N}}
\newcommand{\ulos}{u_\text{L}}
\newcommand{\unlos}{u_\text{N}}
\newcommand{\dlos}{D_\text{L}}
\newcommand{\dnlos}{D_\text{N}}
\newcommand{\slos}{s_\text{L}}
\newcommand{\snlos}{s_\text{N}}
\newcommand{\ilos}{I_\text{L}}
\newcommand{\inlos}{I_\text{N}}

\newcommand{\nuoplos}[1]{\nu_{#1,L}}
\newcommand{\nuopnlos}[1]{\nu_{#1,N}}
\newcommand{\nuoptlos}[1]{\nu_{#1,\tau}}

\newcommand{\imphi}{\tilde{\Phi}}
\newcommand{\imm}{\tilde{\Lambda}}
\newcommand{\immexpr}{\lambda \vol{A}\vol{I}}
\newcommand{\immof}[1]{\tilde{\Lambda}\left(#1\right)}
\newcommand{\imsmm}{\tilde{\mu}^{(2)}}
\newcommand{\imsmmof}[1]{\tilde{\mu}^{(2)}\left(#1\right)}

\begin{document}
\title{Modeling Infrastructure Sharing in mmWave Networks with Shared Spectrum Licenses}
\author{Rebal Jurdi, Abhishek K. Gupta, Jeffrey G. Andrews, and Robert W. Heath, Jr. 
\thanks{Rebal S. Jurdi (rebal@utexas.edu), Jeffrey G. Andrews (jandrews@ece.utexas.edu), and Robert W. Heath, Jr. (rheath@ece.utexas.edu) are with the Wireless Networking and Communications  Group, The University of Texas at Austin, Austin, TX 78712 USA.} 
\thanks{Abhishek K. Gupta (gkrabhi@iitk.ac.in) is with the department of Electrical and Computer Engineering at the Indian Institute of Technology Kanpur, Kanpur, India 208016.}
\thanks{This work was supported in part by Crown Castle International and by the National Science Foundation under Grant No. NSF-CCF-1514275.}}
\maketitle
\begin{abstract}
Competing cellular operators aggressively share infrastructure in many major US markets. If operators also were to share spectrum in next-generation millimeter-wave (mmWave) networks, intra-cellular interference will become correlated with inter-cellular interference. We propose a mathematical framework to model a multi-operator mmWave cellular network with co-located base-stations (BSs). We then characterize the signal-to-interference-plus-noise ratio (SINR) distribution for an arbitrary network and derive its coverage probability. To understand how varying the spatial correlation between different networks affects coverage probability, we derive special results for the two-operator scenario, where we construct the operators' individual networks from a single network via probabilistic coupling. For external validation, we devise a method to quantify and estimate spatial correlation from actual base-station deployments. We compare our two-operator model against an actual macro-cell-dominated network and an actual network primarily comprising distributed-antenna-system (DAS) nodes. Using the actual deployment data to set the parameters of our model, we observe that coverage probabilities for the model and actual deployments not only compare very well to each other, but also match nearly perfectly for the case of the DAS-node-dominated deployment. Another interesting observation is that a network that shares spectrum and infrastructure has a lower rate coverage probability than a network of the same number of BSs that shares neither spectrum nor infrastructure, suggesting that the latter is more suitable for low-rate applications.

\end{abstract}
\section{Introduction}
Millimeter-wave communication will be central to delivering the anticipated performance of next-generation cellular networks \cite{ted,rangan14,rwh5g,zpi}. A key feature of mmWave systems is directional communication \cite{roh,ted13b}, which reduces the effect of out-of-cell interference as compared to communication at UHF frequencies \cite{bai15}, and opens up the possibility of sharing spectrum licenses with no coordination between network operators \cite{boccardi,abi15}. While spectrum sharing is a future possibility, infrastructure sharing is already a reality, and there has been a progression in the cellular operator industry towards sharing network infrastructure such as the network core, backhaul, and cell towers as a means of expanding coverage at a reduced cost \cite{meddour11, fri08}. When multiple closed-access cellular networks share spectrum, inter-network interference adds to intra-network interference. When cellular networks also share cell sites and towers, inter-network interference becomes \emph{coupled} to intra-network interference because many of the BSs of the different networks are stationed at the exact same location. In this paper, we propose a mathematical framework that accurately models the \emph{co-location} of BSs of multiple operators that share spectrum licenses, and suggest how to estimate model parameters from actual deployments.

\subsection{Background}

\begin{figure*}
\centering
\includegraphics[width=\textwidth]{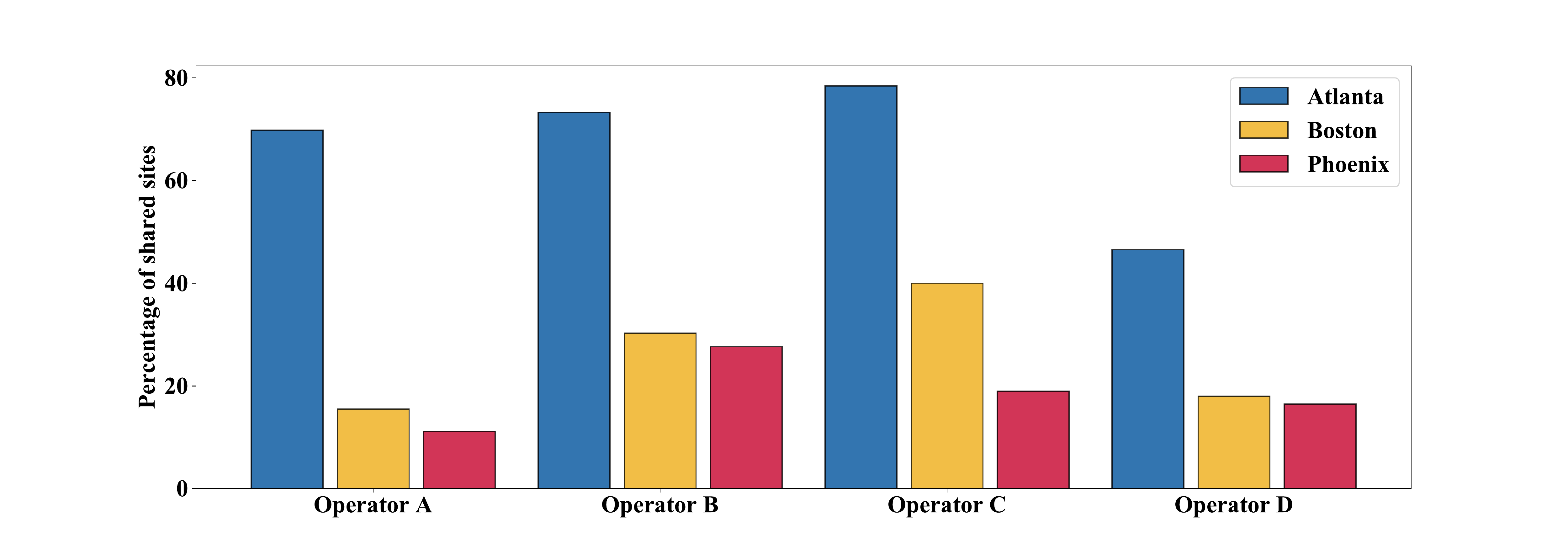}
\caption{The percentage of sites shared \emph{with} one or more cellular operators. Sites considered include macro-cellular towers, rooftops, and DAS nodes. Figures are given for the largest 4 US cellular operators in 3 major cellular market areas (CMAs). In \cmaiii/, \opa/ shares about 11\% of the sites it occupies with competing operators. In \cmai/, three out of four operators share over 66\% of their sites.}
\label{bar_plot}
\end{figure*}

Cellular operators have already been sharing network infrastructure through a variety of business models to increase their coverage and capacity while reducing capital and operational expenditures \cite{khan11}. Infrastructure sharing takes two general forms: passive sharing, i.e. sharing of the space or supporting infrastructure, and active sharing, i.e. sharing of the radio access network (RAN) and network core \cite{gsma12}. Throughout this paper, we will use term ``infrastructure sharing'' synonymously with ``passive sharing''. 
We surveyed three geographically diverse US cellular market areas (CMAs) for instances of passive infrastructure sharing, tallying the sites and \textit{structures} that are occupied by a single operator and those that are shared by two or more operators. We considered 3 CMAs: Atlanta, GA, Boston-Lowell-Brockton-Lawrence-Haverhill, MA-NH, and Phoenix, AZ. Further, we considered the four largest US operators. \figref{bar_plot} shows the percentage of sites shared \emph{with} one or more competing operators. For every market, the bar plot shows the sharing ratio per operator, i.e. the percentage of the operator's BSs that are co-located with those of one or more operators. Sharing ratios range from 11\% to 78\%, with some markets displaying more aggressive site sharing than other markets. This trend is expected to continue in next generation cellular networks through a dense overlay of multi-operator and virtual-host small cells targeting enterprise and entertainment venues, on top of the existing layer of macro towers housing BSs of competing operators \cite{lakis16,okasaka16,ghan13,5gamericas16}.

Spectrum is a valuable asset that can be shared by competing cellular networks \cite{fraidaaa,doyle,kamal09}. Early studies have suggested that cognitive radios are able to efficiently utilize \emph{existing} sparse, sporadically used spectrum and facilitate spectrum sharing between different networks \cite{song12,ian06,kang09}. Spectrum sharing in cellular networks is governed by a variety of licensing policies, known as \textit{authorization regimes}. Licensed access, for instance, permits only the license holder to use their licensed frequency bands, but sharing schemes under this regime remain possible (see \cite{tehrani16} for a detailed taxonomy of spectrum access methods). Co-Primary Shared Access and Licensed Shared Access (LSA) are two access methods that allow license holders, subject to the approval of regulatory authorities, to share some of their spectrum with one another \cite{dahlman2011}, thus jointly bearing license fees and enhancing the utilization of spectrum. Regulatory authorities have recognized the need to reform existing authorization regimes to promote the commercialization of more efficient wireless technologies. The Federal Communication Commission (FCC) and The United Kingdom's Office of Communications (Ofcom) are among many regulatory authorities that are implementing policies to expand access to shared spectrum and actively seeking comments regarding co-existence mechanism for mmWave bands (see \cite{bhattarai16} for a detailed overview of ongoing spectrum sharing initiatives). 

If multiple operators share infrastructure \emph{but not} spectrum, there is no need to model the co-location of their BSs; the transmissions of different networks are orthogonal in frequency, and thus the performance of each of the networks can be studied independently. However, if the operators share both infrastructure \emph{and} spectrum, inter-network interference and intra-network interference become correlated. In this case, it is necessary to model BS co-location to accurately extract the network performance. 

\subsection{Prior Work}
In work related to infrastructure sharing, a statistical approach to model multi-operator networks with shared deployment patterns was presented in \cite{kib15}, but performance of the models therein was evaluated only through simulation. In a subsequent paper by the same authors \cite{kib16}, the impact of spatial clustering, network density, and spectrum access coordination on network coverage in a multi-operator system was studied analytically. In \cite{rebato}, different configurations of infrastructure and spectrum sharing were considered, and corresponding SINR and rate coverage probabilities were compared and evaluated against different channel, antenna, and BS patterns. In \cite{hailu2014}, an adaptive co-primary shared access scheme between co-located RANs that partitions spectrum into private and shared frequency sub-bands was proposed. The economics of infrastructure sharing has also been studied in a game-theoretic framework. In \cite{deng2017}, the relationship between tower companies and cellular operators is examined under different time horizons of the market competition. In \cite{mandrews17}, network pricing and capacity is compared between the case when operators cooperate and when they compete. The papers \cite{deng2017,mandrews17} model pricing and user demand, but offer no framework that enables the coverage analysis of the networks.

In recent work related to mmWave spectrum sharing, the performance of a number of mmWave cellular systems spanning different combinations of spectrum and access sharing methods was analyzed in a stochastic geometry framework \cite{abi15}. Two particular systems were studied: a two-operator system with closed access and full spectrum sharing, and a two-operator system where all the BSs of the two operators are housed on the same towers. Partial BS co-location, however, was not considered in \cite{abi15}. In \cite{abi16}, the feasibility of secondary licensing in licensed mmWave bands is established, yet the model that was used to represent the locations of the primary and secondary BSs did not generalize to scenarios where the primary and secondary networks share infrastructure. In \cite{jpark2017}, the probability of rate coverage of spectrum-shared mmWave networks with inter-operator coordination is determined. In \cite{boccardi2016}, a new authorization system that governs spectrum sharing between multiple operators was introduced under the name of \textit{spectrum pooling}. Preliminary results suggested that spectrum is utilized more efficiently under spectrum pooling than it is under the exclusive spectrum allocation model. Built on this authorization regime, a new mmWave hybrid spectrum access scheme was introduced in \cite{rebato17} that combines an exclusive-access band and another band where spectrum is pooled between multiple operators. 

Prior work such as \cite{abi15,abi16,kib15,kib16,rebato} lacks a model that reproduces any extent of co-location between the BSs of any set of operators, and that allows straightforward analysis of key performance metrics like SINR and rate coverage probabilities.

\subsection{Contributions}



\textbf{Modeling infrastructure sharing between multiple operators:} In this paper, we propose an analytical framework to describe the BS locations of multiple mmWave cellular operators that share infrastructure, i.e. the locations of cell sites that house two operators (sites of dual co-location), sites that house three operators (sites of triple co-location), and so on. Our model is flexible in that it can capture the densities not only of the operators' networks, but also of the sites housing \emph{any} subset of the operators. 

\textbf{Analyzing and comparing the coverage probability of shared networks:} Applying analysis techniques from stochastic geometry, we derive expressions for the SINR probability of coverage of an arbitrary \textit{shared network} in a multi-operator system, i.e. a network that shares infrastructure or spectrum with one or more operators. We then focus on the more tractable case of a system of two networks to understand the effect of varying their spatial correlation on their coverage probabilities. Since varying the spatial correlation of two networks does not necessitate a change in the networks' densities, we consider two perspectives of infrastructure sharing: fixed individual densities (FID), and fixed combined density (FCD). Under FID, the densities of the individual networks' BSs remain constant with varying the the spatial correlation. Under FCD, these densities increase as the correlation increases, but the density of the total BSs remains constant. In practice, FID corresponds to the \emph{relocation} of an operator's BSs to sites that are already occupied by another operator, while FCD corresponds to an operator's \emph{expansion} into such sites. Under these two perspectives, we compare the probability of rate coverage and the median rate between three shared networks undergoing different extents of spatial correlation as well as two \textit{single-operator networks} with different sizes of bandwidth. In this paper, we use the term ``single-operator'' to refer to a network that shares neither infrastructure nor spectrum with any other network and is considered as the baseline. Modeling the correlation between the interference from different operators, which is caused by the co-locations of their BSs, required introducing new analytical techniques such as probabilistic coupling.

\textbf{Estimating model parameters from real deployments:}
To measure how accurately our model reflects the performance of an actual shared network, we compare the SINR probability of coverage obtained for our model and for actual networks. Since mmWave deployments currently do not exist, we suggest ways of extracting model parameters from actual deployments, namely, the densities of the shared networks and their spatial correlation. We consider both macro-tower-dominated deployments and deployments predominantly comprised of distributed antenna system nodes in major US cellular markets. Our results show that coverage probabilities for the PPP model and actual deployments compare very well, and they are even almost identical in the case of the DAS-node-dominated deployment.

The rest of the paper is organized as follows. Section~\ref{sys} describes the channel and multi-operator system model. Section~\ref{cov} gives the expressions of the probability of SINR and rate coverage. In Section~\ref{dual}, we consider the two-operator model from a different angle which allows us to quantify the spatial correlation between two BS deployments, and estimate its value from actual deployments. Section~\ref{num} presents numerical results and provides some insights. Finally, we conclude in Section~\ref{conc}.

\section{System Model}\label{sys}
In this section, we describe how to represent the BS locations of $M$ mmWave cellular operators that share infrastructure and spectrum licenses. We construct the operators' networks by combining independent point processes to produce any amount of co-location of any subset of the $M$ networks. Our model reduces to the two extremes of co-location described in \cite{abi15}: \textit{full independence}, where BSs of each operator are represented by their own point process, and \textit{full overlap}, where BSs of all operators are at the exact same locations and thus represented by a \emph{single} point process. Before we proceed to the mathematical description of our model, we introduce a few variables. Let $\op=\set{1,2,\dots,M}$ be the set of operators and $\peeof{\op}$ the power set of $\op$. Let $\set{\Phi_\ess}$ be a collection of \emph{independent}, homogeneous PPPs, where the index $\ess\in\peeof{\op}$.

We use point processes from the collection $\set{\Phi_\ess}_{\ess\in\peeof{\op}}$ to build the operators' networks, and we regard these point processes as the basic building blocks of our model; hence, we refer to them as \textit{blocks}. Every block $\Phi_\ess$ represents the (random) locations of the cell sites that are shared by the elements of $\ess$. For example, set $M=2$, then $\Phi_{\set{1,2}}$ becomes the PPP that describes the sites housing the BSs of both $\netw$ and $\nett$, i.e. sites of dual co-location. For a general value of $M$ and an arbitrary subset $\ess$, $\Phi_\ess$ could represent sites of dual co-location, triple co-location, $\ldots$ , and $M$-tuple co-location. Choosing a subset $\ess$ and assigning a density $\lambda_\ess$ to $\Phi_\ess$ is equivalent to saying that the density of the cell sites housing the members of $\ess$, and the members of $\ess$ \emph{only}, is $\lambda_\ess$. Consider, for example, that Operator $m$ is a member of not only $\ess$, but $\ess'$ as well. Consequently, Network $m$ contains the cell sites described by $\Phi_\ess$ and $\Phi_{\ess'}$, or equivalently by $\Phi_\ess\cup\Phi_{\ess'}$. Exhausting all acts of co-location with every possible un-ordered tuple $\ess$ of operators, Network $m$ would contain \emph{exactly} the cell sites described by the point process $\Phi_m$ which is given as
\begin{equation}\label{eq_orthogonalized}
\Phi_m = \bigcup_{{\ess:\;m\in\ess}} \Phi_\ess,
\end{equation}
and has a density 
\begin{equation}\label{eq_orthogonalized_density}
\lambda_m = \sum_{{\ess:\;m\in\ess}} \lambda_\ess.
\end{equation}
We shall refer to $\Phi_m$ as the \emph{individual} point process of Operator $m$. An important property of $\Phi_m$ is that it is a PPP since it is the superposition of independent PPPs \cite{baccelli}. Additionally, we have a probabilistic guarantee that the construction of $\Phi_m$ avoids ``double-counting'' of sites, i.e. for any $\ess,\,\ess'\in\peeof{\op}$, $\Phi_\ess\cap\Phi_{\ess'}=\emptyset$, almost surely. This is true since any collection of independent PPPs has no points in common, almost surely.

It is important to realize the difference between $\set{\Phi_\ess}_{\ess\in\peeof{\op}}$, and $\set{\Phi_m}_{m\in\op}$. The first is a collection of independent PPPs that are the fundamental building blocks of the model. The second is a collection of the individual PPPs that characterize the operators' BS locations and are the outcome of combining different blocks. Consider again the two-operator example, i.e. when $M=2$. There are three blocks, $\Phi_{\set{1}}$, $\Phi_{\set{2}}$ and $\Phi_{\set{1,2}}$, and two individual point processes, $\Phi_1=\Phi_{\set{1}} \; \cup \; \Phi_{\set{1,2}}$ and  $\Phi_2=\Phi_{\set{2}} \; \cup \; \Phi_{\set{1,2}}$.


As for the spectrum sharing model, we make the simplifying assumption that all operators own licenses of an equal amount of spectrum, and that these licenses are shared. We now make the following assumptions about the blockage and channel models.
\paragraph*{Blocking model} We assume the independent blocking model where the link established between the typical user and a BS located at a distance $r$ away can either be \textit{line-of-sight} (LOS), denoted by $\text{L}$, with probability $\plos(r)$ or \textit{non-line-of-sight} (NLOS), denoted by $\text{N}$, with a probability $\pnlos(r)=1-\plos(r)$. We adopt the exponential blocking model introduced in \cite{bai15}, where $\plos(r)=\exp(-\beta r)$. Hence, conditioned on the typical user, each system block $\Phi_\ess$ of density $\lambda_\ess$ is divided into two independent non-homogeneous PPP as a direct result of the independent thinning theorem \cite{baccelli}, and we obtain the two \textit{sub-blocks}:
\begin{itemize}
\item $\los_\ess$ containing all BSs with LOS links to the user. It has density $\lambda_{\ess,\text{L}}(r)=\plos(r)\lambda_\ess$ and measure $\Lambda_{\ess,\text{L}}\,$.

\item $\nlos_\ess$ containing all BS with NLOS links. It has density $\lambda_{\ess,\text{N}}(r)=\pnlos(r)\lambda_\ess$ and measure $\Lambda_{\ess,\text{N}}\,$.
\end{itemize}
Then, it follows that the average number of BSs in the sub-blocks $\los_\ess$ and $\nlos_\ess$ in the Euclidean ball $B_0(r)$ centered at the origin and of radius $r$ is
\begin{align}
\Lambda_{\ess,\text{L}}(B_0(r)) &= 2\pi\lambda_\ess \int_0^r \plos(t)t\diff t = \frac{2\pi\lambda_\ess}{\beta^2}\gamma(2,\beta r), \\
\Lambda_{\ess,\text{N}}(B_0(r)) &= 2\pi\lambda_\ess \int_0^r \pnlos(t)t\diff t = \pi\lambda_\ess\left( r^2 - \frac{2}{\beta^2}\gamma(2,\beta r) \right),
\end{align}
where $\gamma(.,.)$ is the lower incomplete gamma function.

\paragraph*{Transmit and noise power} We assume that all BSs transmit at a fixed power $P_\text{t}$. We consider a noise power spectral density $N_0$ and a total bandwidth $B$. 

\paragraph*{Path loss} We consider the power-law path loss functions for LOS and NLOS links:
\begin{align*}
\ell_\text{L}(r) = \clos r^{-\alos}, \text{ and } \ell_\text{N}(r) = \cnlos r^{-\anlos},
\end{align*}
where $\clos$ and $\cnlos$ correspond to the power attenuation at $r=1$ for LOS and NLOS links.

\paragraph*{Directivity gain} Similar to \cite{bai15}, base stations are equipped with steerable antennas characterized by a main-lobe gain $\sfgmax$ and side-lobe gain $\sfgmin$. Even though users will also have directional antennas, the analysis would be equivalent to the case of aggregating the transmitter and receiver gains at BS antennas. Therefore, we assume that user mobile devices have a single omni-directional antenna as in \cite{abi16}, and that all points $\set{X_k}$ representing BS locations are endowed by marks $\set{G_k}$ which are IID Bernoulli distributed with PMF
\begin{equation}
G_k=\begin{cases}
\sfgmax &\text{w.p.}\ \theta_b/\pi\\
\sfgmin &\text{w.p.}\ (\pi-\theta_b)/\pi,
\end{cases}
\end{equation} 
where $\theta_b$ is the half beamwidth and assumed to be identical accross all BSs. Since signals received from co-located BSs are transmitted from antenna arrays pointed in different directions to serve different users, we can assume that directionality gains are independent. In reality, actual array patterns can be different from those produced by this model because of scattering and dispersion \cite{valenzuela}. Nevertheless, we use this model for analytical tractability.

\paragraph*{Association rule} We consider a closed-access system where the users can only connect to the base stations of their parent network. Moreover, the typical user associates to the BS that corresponds to the smallest path loss, or equivalently, the BS providing the maximum received signal averaged over fading. Once the BS is chosen and a link is established, the BS antenna array aligns its beam with the user to ensure maximum signal gain. The typical user could form either a LOS or a NLOS link with the serving station. 

\paragraph*{Small-scale fading} We assume that the channel undergoes flat Rayleigh fading. Equivalently, the fading power $H_k$ of the signal received from the BS at $X_k$ is exponentially distributed with unit mean. We verify in~\secref{num} that the relative performance remains unchanged when Nakagami fading and lognormal shadowing are used. Despite the fact that the large-scale propagation losses of co-located transmitters of opposite networks are equal at any distance, we assume that signals received from these transmitters at any point undergo independent fades. This is reasonable given the different locations of BS antennas on the tower are typically further than the (vertical) coherence distance of the channel.

\section{Coverage Analysis}\label{cov}
\renewcommand{\arraystretch}{1.5} 

We use the SINR probability of coverage as the system performance metric, which is defined as the value of the SINR complementary cumulative distribution function (CCDF) at a threshold $T$
\begin{align}\label{pcov}
\pcovw{T} = \prob{\mathrm{\sfsinr}>T}.
\end{align}
Suppose that the typical user associates with $b^\text{th}$ BS of the $n^\text{th}$ network at a distance $R$ via a link of type $\tauof{b}{n}$ which can be LOS or NLOS. We define $c_{\tauof{b}{n}}$ and $\alpha_{\tauof{b}{n}}$ to be the path loss constant and exponent corresponding to $\tauof{b}{n}$, and $\sigma^2$ to be the thermal noise power normalized by the transmit power, i.e. $\sigma^2=N_0 B/P_\text{t}$. We also define $I$ to be the interference from all blocks and is expressed as
\begin{align}\label{interference1}
\begin{split}
	I &= \sum_{\substack{i:\,X_i\in \Phi_n\\i\neq b}} c_{\tauof{i}{n}} H_{i,n}G_{i,n}||X_{i,n}||^{-\alpha_{\tauof{i}{n}}} \\
	&+ \sum_{\substack{m\in O\\m\neq n}} \sum_{j:\,X_j\in \Phi_m} c_{\tauof{j}{m}} H_{j,m}G_{j,m}||X_{j,m}||^{-\alpha_{\tauof{j}{m}}}.
\end{split}
\end{align}

Therefore, the SINR of the typical user is
\begin{align}
\begin{split}\label{sinr}
\sfsinr&=\displaystyle\frac{c_{\tauof{b}{n}} H_{b,n}\sfgmax R^{-\alpha_{\tauof{b}{n}}}}{\sigma^2 + I}.
\end{split}
\end{align}
The first term of the sum accounts for the interference from BSs of the same operator, while the second term describes the interference from all BSs of different operators. Note that $X_{j,m}$ and $X_{j',m'}$, the locations of BS $j$ of network $m$ and BS $j'$ of network $m'$, need not be distinct. If $m,m'\in\ess$, then $\Phi_m$ and $\Phi_{m'}$ share points in common, as they are both derived from block $\Phi_\ess$.

In the remainder of this section, we analyze the coverage probability of a typical user of $\netw$ since the coverage analysis of all networks is mathematically identical. The networks could have different coverage due to the various parameter values. However, this does change the \emph{analysis}. We first investigate the association of the typical user of $\netw$ to any of its BSs. We then compute its SINR probability of coverage and derive the rate probability of coverage which is a tangible metric in quantifying user experience.
\begin{table*}[t!]
	\centering
	\caption{Summary of notation.}
	\label{tab:SummaryOfNotations}
		\begin{tabularx}{\textwidth}{|c|X|} 
			\hline
			\textbf{Notation} 								& \textbf{Description} \\
			\hline
			$\op,\;\peeof{\op}$										&	Set of all cellular operators and its power set		\\
			\hline
			$\ess,\;\essp,\;\tee$							&	Subsets of the power set of all cellular operators		\\
			\hline
			$\Phi_\ess$, $\ess\in \peeof{\op}$			&	Point process describing locations of structures (e.g., towers, rooftops) occupied by operators in $\ess$		\\
			\hline
			$\Phi_m$, $m\in \mathbb{Z}_+$				&	Point process describing locations of structures occupied by operator $m$		\\
			\hline
			$\lambda_m\;,\lambda_m'$					& Densities of $\Phi_m$ and $\Phi'_m=\Phi_m\setminus \bigcup_{n\notin\ess}\Phi_n$ \\
			\hline
			$\los_\ess,\;\nlos_\ess,\;\tlos_\ess$			&	LOS, NLOS, and arbitrary sub-brackets of $\Phi_\ess$ as seen from a user at the origin		\\
			\hline
			$\lambda_\ess,\;\Lambda_\ess$			&	Density of $\Phi_\ess$ and its mean measure 	\\
			\hline
			$\text{L},\;\text{N},\;\tau$										& Subscripts used to denote a LOS, NLOS, or unspecified type of link \\
			\hline
			$\nuoptlos{\ess}(r),\;\nuoptlos{m}(r)$  & Mean measures of the Euclidean ball of radius $r$ centered around the origin for the point processes/brackets $\tlos_{\ess}$ and $\tlos_m$, respectively \\
			\hline
			$\plos(r),\;\pnlos(r)$									& Probability that the user establishes a link with a LOS, NLOS BS given the length of the link is $r$ \\
			\hline
			$\sfgmax,\sfgmin,\theta_b$			&	Maximum and minimum antenna array gains, and half beamwidth		\\
			\hline
			$\prob{\cdot}$										&	Probability of an event			\\
			\hline
			$\pcovw{\cdot},\;\pcovt{\cdot}{\tee}$			& Probability of SINR coverage, and the same probability conditioned on associating with bracket/sub-bracket $\tee$ \\
			\hline
			$\expect{X},\;\expectw{Y}{X}$			& Expectation of $X$ and expectation of $X$ taken with respect to the distribution of $Y$ \\
			\hline
			$\lapof{X}{\cdot},\;\lapof{X|Y}{\cdot}$	  &	Laplace transform and conditional Laplace transform	\\
			\hline
			$\lapvar{\cdot\,;\,\tlos}$				& Laplace transform given association to bracket or sub-bracket $\tlos$ \\
			\hline
		\end{tabularx}
\end{table*}

\subsection{Association Criterion}
The probability that a typical user of $\netw$ is covered depends on what block they are associated with. Association could take place through any of the $2^{M-1}$ blocks of $\set{\Phi_{\ess}}$, $1\in\ess$, and any of their sub-blocks. Since these blocks and sub-blocks are independent, the events of associating to distinct blocks are disjoint. Hence, we can compute the \emph{total} probability of coverage by adding the \emph{joint} probabilities of coverage \emph{and} association.

We first define some notation. Let $\nu_{\ess,\tau}=\Lambda_{\ess,\tau} \circ B_0$, where $\Lambda_{\ess,\tau}$ is the measure for the appropriate sub-block of $\tlos_\ess$, and the subscript $\tau$ denotes an arbitrary link type. The operator $\circ$ denotes \textit{composition}, i.e., $\nu_{\ess,\tau}(r)=\Lambda_{\ess,\tau} \left(B_0(r)\right)$. Additionally, let $\dlos$ be the \textit{exclusion function} of LOS transmitters of $\netw$ when the user is associated with a NLOS transmitter of the same network. Similarly, let $\dnlos$ be the exclusion function of NLOS transmitters when the user is associated with a LOS transmitter. An exclusion function gives the radius of the region around the tagged BS within which no other BSs in the same or different blocks exist. These functions are given in \cite{bai15} as 
\begin{align*}
\dlos(r)=\left(\frac{\clos}{\cnlos}\right)^\frac{1}{\alos} r^\frac{\anlos}{\alos},\text{ and } \dnlos(r)=\left(\frac{\cnlos}{\clos}\right)^\frac{1}{\anlos} r^\frac{\alos}{\anlos}.
\end{align*}
Moreover, let $\assoc_{\tlos_\ess}$ be the event of association with sub-block $\tlos_\ess$. Define $\pcovt{T}{\Phi_\ess}$ and $\pcovt{T}{\tlos_\ess}$ to be the probabilities that the user is in coverage for a threshold $T$ \emph{and} that the user is associated to a BS in any block $\Phi_\ess$ and sub-block $\tlos_\ess$ thereof:
\begin{equation}
\pcovt{T}{\tlos_\ess} = \prob{\sfsinr > T \;\medcap\;\assoc_{\tlos_\ess}},
\end{equation}
These probabilities can be obtained by integrating the CCDF of the appropriate SINR given by \eqref{sinr},~\eqref{interference1}, and weighted by the probability density function (PDF) $\fr{\cdot}{\tlos_\ess}$ of the length $R$ of the established link with a BS of $\tlos_\ess$ as
\begin{align}\label{eq_pcov_tier}
\begin{split}
&\pcovt{T}{\tlos_\ess} = \\
&\int_{r\geq 0} \prob{\sfsinr>T\;\medcap\;\assoc_{\tlos_\ess}|\,R=r} \fr{r}{\tlos_\ess} \diff r.
\end{split}
\end{align}
What remains to be derived is the PDF $\fr{.}{\tlos_{\ess}}$ for an arbitrary sub-block $\tlos_{\ess}$. To accomplish this, we follow the derivation in \cite{abi15}. We draw an analogy between the sub-blocks that a typical user of $\netw$ can associate with, defined in this paper, and the \textit{tiers} as defined in \cite{abi15}. In \cite{abi15}, a typical user (of $\netw$, let's say) is permitted to access $\netw$, their home network, and every other network that is in the access class of $\netw$. In a closed-access system, the access class is $\netw$, and in an open-access system, the access class is all networks. The key to computing $\fr{r}{\tlos_{\ess}}$ is computing the probability $\fro{r}{\tlos_{\ess}}$ that all BSs of every other sub-block $\tlos_\tee,\,\tee\in\peeof{\op}\setminus\ess,\,1\in\ess$ that are accessible by the typical user are outside the exclusion radius $r$. For an arbitrary sub-block $\tlos$, this is given by the void probability $\pvoid{\tlos}{r}=\prob{\tlos(\ball{r})=0}$. Since all sub-blocks are mutually independent, $\fro{r}{\los_{\ess}}$ is given as the product of void probabilities
\begin{align}
\begin{split}
\fro{r}{\los_{\ess}} 
&= \pvoid{\nlos_{\ess}}{\dlos(r)} \\
&\cdot
\prod_{\substack{\tee\in\peeof{\op}\setminus\ess\\ 1\in\tee}}
\pvoid{\los_{\tee}}{r}\,\pvoid{\nlos_{\tee}}{\dlos(r)}.
\end{split}
\end{align}

The density $\fr{\cdot}{\los}$ is obtained according to \cite[Section~V-C]{sgprimer} and \cite[Equation~(9)]{abi15} as
\begin{align}
\begin{split}
\fr{r}{\los_\ess} &= \frac{\diff}{\diff r}\left(\pvoid{\los_\ess}{r}\right) \fro{r}{\los_{\ess}} \\
&= 2\pi\lambda_\ess\, r\,\plos(r) e^{-\nuoplos{\ess}(r)-\nuopnlos{\ess}(\dlos(r))} \\
&\cdot
\prod_{\substack{\tee\in\peeof{\op}\setminus\ess\\ 1\in\tee}} 
e^{-\nuoplos{\tee}(r)-\nuopnlos{\tee}(\dlos(r))}.
\end{split}
\end{align}

Since the events $\set{\assoc_\tlos}$ of association with different sub-blocks $\set{\tlos}$ are disjoint \cite{abi15}, the SINR coverage probability $\pcovw{T}$ for the typical user is obtained by adding these individual block coverage probabilities over all accessible block of $\netw$:
\begin{align}\label{total_prob}
\begin{split}
\pcovw{T}	&= 
\sum_{\substack{\ess\in\peeof{O}\\ 1\in\ess}} \pcovt{T}{\Phi_\ess} \\
&= \sum_{\substack{\ess\in\peeof{O}\\ 1\in\ess}} \pcovt{T}{\los_\ess} + \pcovt{T}{\nlos_\ess}.
\end{split}
\end{align}
Next, we derive the expression for $\pcovt{T}{\tlos_\ess}$ for an arbitrary sub-block.

\subsection{Interference Characterization}
Computing the probability of coverage under Rayleigh fading can be readily reduced to finding the Laplace transform of interference. To illustrate this, \eqref{pcov} can be expanded as
\begin{align}\label{eq_pcov2}
\begin{split}
\prob{\sfsinr>T \,|\, R} &= \prob{c_\tau \sfgmax H R^{-\alpha_\tau} > T\left(\sigma^2+I\right) \,\bigg{|}\, R} \\
				 &= \expect{\overbar{F}_{H}\left(\left(\frac{R^{\alpha_\tau}T}{c_\tau \sfgmax }\right)(\sigma^2+I)  \,\bigg{|}\, R,\,I\right)},
\end{split}
\end{align}
where $\overbar{F}_{H}$ is the CCDF of the fading encountered by the signal emitted from the tagged BS which can be expressed as a single exponential. Noting that $\overbar{F}_{H}(u)=e^{-u}$, \eqref{eq_pcov2} becomes
\begin{align}\label{eq_pcov_final}
\begin{split}
\prob{\sfsinr>T \,|\, R} &= \exp{\left(-\frac{\sigma^2 R^{\alpha_\tau}T}{c_\tau \sfgmax} \right)} \lap{I|R} \left( {\frac{R^{\alpha_\tau}T}{c_\tau \sfgmax }} \right).
\end{split}
\end{align}
Since the BSs of $\netw$ are spatially co-located with those of other networks, inter-network interference is no longer independent and $\mathcal{L}_{I|R}$ is not  the product of Laplace transforms of inter-network and intra-network interference. To resolve this, we reformulate the interference expression in \eqref{interference1}, which is given with respect to the correlated point processes of $\set{\Phi_m}_{m\in O}$, as a sum over the uncorrelated block of $\set{\Phi_\ess}_{\ess\in\peeof{O}}$. We give the resulting expression in the following proposition.

\begin{proposition}\label{prop_interference_laplace}
Given that the typical user associates to a BS at a distance $r$ in the LOS sub-block $\los_\tee$, $\tee\in\peeof{O}$, then the Laplace transform of the interference random variable $\lap{I_\text{L}}(s)$ is given as
\begin{flalign}
\begin{split}\label{eq_interference_laplace}
&\ulos(s,r) ^{|\tee|-1} \\
\cdot&\prod_{\ess:\,1\notin\ess}   \exp\left(-2\pi\lambda_\ess \int_{t\geq 0} \left(1-\ulos(s,t)^{|\ess|}\right)  \plos(t)t\diff t \right) \\
\cdot&\prod_{\essp:\,1\in\essp} \exp\left(-2\pi\lambda_\essp \int_{t\geq r} \left(1-\ulos(s,t)^{|\essp|}\right)  \plos(t)t\diff t  \right) \\
\cdot&\prod_{\ess:\,1\notin\ess}   \exp\left(-2\pi\lambda_\ess \int_{t\geq 0} \left(1-\unlos(s,t)^{|\ess|}\right)  \pnlos(t)t\diff t \right) \\
\cdot&\prod_{\essp:\,1\in\essp} \exp\left(-2\pi\lambda_\essp \int_{t\geq D_L(r)} \left(1-\unlos(s,t)^{|\essp|}\right)  \pnlos(t)t\diff t  \right),
\end{split}
\end{flalign}
where $u_\tau(s,t)=\expectw{G}{\lapof{H|G}{sc_\tau G t^{-a_\tau}}}$, and $G$ the antenna gain random variable. Moreover, if fading power is exponentially distributed with unit mean and antenna gain follows a Bernoulli distribution, $\lap{I_L}(s)$ is given as
\begin{align}\label{eq:utau}
u_\tau(s,t)&=\frac{\theta_b/\pi}{1+sc_\tau\sfgmax t^{-a_\tau}}+\frac{(\pi-\theta_b)/\pi}{1+sc_\tau\sfgmin t^{-a_\tau}}.
\end{align}
\end{proposition}
\begin{IEEEproof}
The proof is detailed in Appendix \ref{app_a}.
\end{IEEEproof}
The Laplace transform is the product of Laplace transforms of independent random variables corresponding to different classes of BSs, grouped according to the network they belong to as well as the type of potential link they can establish with the user.  The first term of \eqref{eq_interference_laplace} represents the contribution of the LOS BSs of different operators co-located with the typical user's tagged BS at a distance $r$ away, since there are a total of $|\tee|$ co-located BSs in sub-block $\los_\tee$. The second term gives the LOS interference from all the sites where no BS of $\netw$ is deployed. The third term gives the LOS interference from all the sites housing BSs of $\netw$ averaged outside the exclusion ball associated with the tagged BS. The remaining terms are almost identical to the second and third terms, with the only difference being that they account for NLOS interference. 

The probability of coverage is expressed in terms of the interference Laplace transform (see~\eqref{eq_pcov_tier} and~\eqref{eq_pcov_final}). Now that we have determined the Laplace transform, we give the coverage probability in the next proposition. 
\begin{corollary}\label{prop_pcov_multi}
The SINR probability of coverage of a typical user of $\netw$ in a multi-operator system is given by
\begin{align}\label{eq_pcov_multi}
\begin{split}
\pcovw{T} &= \int_{r\geq 0} \sum_{\substack{\ess\in\peeof{O}\\ 1\in\ess}} \Big{(} e^{-\sigma^2 \slos}\lapvarof{I\,;\,\los_\ess}{\slos} \fr{r}{\los_\ess}  \\
&+\, e^{-\sigma^2 \snlos}\lapvarof{I\,;\,\nlos_\ess}{\snlos} \fr{r}{\nlos_\ess} \Big{)} \diff r,
\end{split}
\end{align}
where $\lapvar{I\,;\,\tlos}$ is the Laplace transform of interference given the event $\assoc_\tlos$, and $\slos$ and $\snlos$ are given by
\begin{equation*}
\slos=\frac{Tr^\alos}{\clos\sfgmax},\;\snlos=\frac{Tr^\anlos}{\cnlos\sfgmax}
\end{equation*}
\end{corollary}
\begin{IEEEproof}
The result follows by substituting the expression of the Laplace transform of the interference random variable given from Proposition~\ref{prop_interference_laplace} in \eqref{eq_pcov_final}.
\end{IEEEproof}

The ultimate metric for evaluating the performance of a cellular network is the per-user downlink rate distribution since it reflects an aspect of service quality experienced by the user. We can transform the SINR coverage probability into a rate coverage probability with a few assumptions. The amount of bandwidth resources allotted to a typical user is a function of the total number of users served by the associated BS, as well as the total available bandwidth $B$. We assume a fair resource allocation algorithm where the BS scheduler divides bandwidth resources equally among each of the $N_U$ users of spatial density $\lambda_U$. Due to the closed-access nature of our multi-operator system, users of a particular operator can only connect to their operator's home network. Hence, the mean number of connected users in a cell can be given based on the approximate load model in \cite{abi15, singh13, singh15} as $N_U=1+1.28\left(\frac{\lambda_U}{\lambda}\right)$. Finally, the probability that a typical user of $\netw$ experiences a rate of at least $R$ bps is $\prob{\sfrate>R} =  \pcovw{2^{R N_U/B}-1}$.

\section{The Two-Operator Case}\label{dual}
In this section, we analyze the probability of coverage for the two-operator case. This scenario is an important special case because it allows to parameterize the system using only three quantities: the densities of the two networks and the extent of overlap between them. In addition, it is difficult to simplify the general case because one has to sweep the densities $\set{\lambda_\ess}$ of all the underlying blocks $\set{\Phi_\ess}$, and consider all possible ways these blocks could be combined to construct the point processes $\set{\Phi_m}$ describing the BS locations of every operator.

In this section, we describe how to construct the operator point processes when there are only two operators, and analyze the probability of coverage for $\netw$. Next, we describe how to estimate the parameters of our model from actual two-operator deployments. Then, we consider two perspectives to study how the coverage probability changes as a function of the overlap between the two networks.

\subsection{Two-Operator Model}\label{dual_model}
The two-operator cases allows for a more natural construction of the operator point processes that allows to describe the model using a few parameters: the densities of the two networks and the extent of overlap between them. We start with a mother point process $\Phi$ of density $\lambda$, and \emph{extract} the two child point processes from it. We identify $\Phi$ as sites managed by a network infrastructure provider, and $\Phi_1$ and $\Phi_2$ as sites leased to two independent operators. We build our model by capturing the density of the first network ($\netw$) and the density of the second ($\nett$), denoted by $\lambda_1$ and $\lambda_2$. We introduce the \textit{overlap coefficient} $\rho(\cdot,\,\cdot)$ as a measure of spatial correlation between $\netw$ and $\nett$ over two sets. For any two given sets $A$ and $B$, $\rho(A,\,B)$ is a function of the covariance between the number of $\netw$ sites in $A$ and that of $\nett$ in $B$ as
\begin{equation}\label{overlap_coef}
\rho(A,\, B) \triangleq \frac{\cov{\Phi_1(A)}{\Phi_2(B)}}{\expect{\Phi(A\cap B)}}.
\end{equation}
Notice that the numerator of the overlap coefficient is a function of the \textit{cross-moment} $\expect{\Phi_1(A)\Phi_2(B)}$ which describes the interaction between two point processes $\Phi_1$ and $\Phi_2$. The normalization by the total density $\lambda$ is necessary so that one can compare the spatial correlation of two networks across different markets of distinct sizes. Since we are interested in the correlation of these two point processes over the entire geographical window $W$, we use $\rho$ as a shorthand notation to $\rho(W\,W)$. Proposition \ref{dual_construction} shows that $\rho$ is in fact directly proportional to $\lambda$, which is a direct result of extracting the child processes from a mother process that is Poisson. If the mother process is not Poisson, the overlap coefficient will not necessarily be proportional to the total density $\lambda$.

We now explain how to mathematically construct $\Phi_1$ and $\Phi_2$ from $\Phi$. The key to construct $\Phi_1$ and $\Phi_2$ from $\Phi$ is the \emph{coupling} technique, where we enforce that the derived point processes have some points $\set{X_k}$ of $\Phi$ in common by coupling them on the same probability space. We begin by marking the points of $\Phi$ with independent random variables $\set{U_k}_{k\geq 0}$ uniformly distributed between 0 and 1. We next consider two parameters $a$ and $b$, where $0\leq b\leq a\leq 1$, and the \textit{retention functions} $q_1(X_k)=a$ and $q_2(X_k)=1-b$. A retention function assigns to every point of a point process a probability of being retained, or alternatively, discarded \cite{baccelli}. Here the probability that a point $X_k$ in $\Phi$ is retained by $\Phi_1$ and $\Phi_2$ is $a$ and $1-b$, respectively. As a result of thinning $\Phi$ separately with $q_1$ and $q_2$, the following two child processes can be obtained
\begin{align*}
\Phi_1(\omega) &=  \set{ X_k(\omega) \mid U_k(\omega)\leq a },\\
\Phi_2(\omega) &= \set{ X_k(\omega) \mid U_k(\omega) > b },
\end{align*}
where $\omega\in\Omega$, and $\Omega$ is the common sample space. Let $\Phi_{12}=\Phi_{\set{1,2}}=\Phi_1\cap\Phi_2$ be the point process describing the locations of shared sites, and let $\lambda_{12}$ be its density. We now give a proposition that validates the above construction of the individual networks from a greater one.

\begin{proposition}\label{dual_construction}
Given $\lambda_1$, $\lambda_2$ and an overlap coefficient $\rho$ on a common geographical window $W\in\mathbb{R}^2$, the thinning of $\Phi$ with $a=\frac{\lambda_1}{\lambda}$ and $b=1-\frac{\lambda_2}{\lambda}$ yields $\Phi_1$ with density $\lambda_1$,  $\Phi_2$ with density $\lambda_2$, and $\lambda=\frac{\lambda_{12}}{\rho}$. Furthermore, $\Phi_1$, $\Phi_2$, and $\Phi_{12}$ are PPPs.
\end{proposition}
\begin{IEEEproof}
See Appendix \ref{app_c}.
\end{IEEEproof}
Since $\rho$ turns out to be the fraction of co-located BSs, and $0\leq\rho\leq 1$, we use it as a proxy for $\lambda_{12}$ to obtain $\lambda_1 + \lambda_2 -\rho\lambda = \lambda$. In the next proposition, we give the expression for the interference Laplace transform, which is the stepping stone towards the coverage probability expression.

\begin{proposition}
Given that the typical user establishes a LOS link of length $r$ with a $BS$ of $\netw$ that is not co-located with a BS of $\nett$, the Laplace transform of the interference random variable $\lap{I_\text{L}}(s)$ is
\begin{align}\label{laplos_dual_nocoloc}
\begin{split}
&\exp\left(-\cst\int\nolimits_0^{r} \left(1-\ulos(s,t)\right)(1-a)\plos(t)t\, \diff t \right) \\
\cdot &\exp\left(-\cst\int\nolimits_r^{+\infty} \left(1-\ulos(s,t)\right)\left(1+\rho \ulos(s,t)\right)\plos(t)t\, \diff t \right) \\
\cdot &\exp\left(-\cst\int\nolimits_0^{\dlos(r)} \left(1-\unlos(s,t)\right)(1-a)\pnlos(t)t\, \diff t \right) \\
\cdot &\exp\left(-\cst\int\nolimits_{\dlos(r)}^{+\infty} \left(1-\unlos(s,t)\right)\left(1+\rho \unlos(s,t)\right)\pnlos(t)t\, \diff t\right).
\end{split}
\end{align}

If, otherwise, the tagged BS of $\netw$ is co-located with a BS of $\nett$, then the Laplace transform of interference is given as
\begin{align}\label{laplos_dual_coloc}
\lap{I_\text{L}}'(s) = \ulos(s,r) \, \lap{I_\text{L}}(s).
\end{align}
\end{proposition}
\begin{IEEEproof}
This follows directly from Proposition \ref{prop_interference_laplace} with $M=2$, $\ess\in\set{\set{1},\,\set{1,2}}$, and $\essp=\set{2}$. 
\end{IEEEproof}
We examine the extreme cases of infrastructure sharing, namely full spatial independence and full spatial co-location, through \eqref{laplos_dual_nocoloc}.  These two cases mimic the \textit{closed access with full spectrum sharing} and \textit{co-located BSs with closed access and full spectrum license sharing} systems studied in \cite{abi15}. In particular, values of $\rho=1$ (and $a=1$) fold the Laplace transform expression in \eqref{laplos_dual_nocoloc} into a product of the second and the fourth terms. These two factors account for the interference contribution from all BSs of the two networks that are located in the LOS and NLOS exclusion balls $B_0(r)$ and $B_0(\dlos(r))$ centered at the typical user. Note that the Laplace transform expression for NLOS interference follows the same derivation, so it is excluded. 

Before giving the final proposition, we introduce some notation. Let $\Phi_1'=\Phi_1\setminus\Phi_2$ and $\Phi_2'=\Phi_2\setminus\Phi_1$ be two point processes with respective densities $\lambda_1'$ and $\lambda_2'$. Now, we give a corollary that relates the probability of coverage of a typical subscriber of $\netw$ in a two-operator system to the different system parameters. 

\begin{corollary}
The SINR probability of coverage of a typical user of $\netw$ in a two-operator system is given by
\begin{align}\label{eq_pcov_dual}
\begin{split}
\pcovw{T} &= \int\limits_{r\geq 0} \sum_{\ess} \Big{(} e^{-\sigma^2 s_L}\lapvarof{I\,;\,\los_\ess}{\slos} \fr{r}{\los_\ess}  \\
&+\, e^{-\sigma^2 \snlos}\lapvarof{I\,;\,\nlos_\ess}{\snlos} \fr{r}{\nlos_\ess} \Big{)} \diff r,
\end{split}
\end{align}
where $\slos$ and $\snlos$ have the same definition as in Corollary~\ref{prop_pcov_multi}.
\end{corollary}
\begin{IEEEproof}
This is a special case of Proposition~\ref{prop_pcov_multi} with $M=2$ and $\ess\in\set{\set{1},\,\set{1,2}}$.
\end{IEEEproof}
As in Corollary~\ref{prop_pcov_multi}, the inner summation in ~\eqref{eq_pcov_dual} is across all collections $\ess$ of operators in which operator 1 is present. Every block $\Phi_\ess$ is divided into two sub-blocks $\los_\ess$ and $\nlos_\ess$, and their contributions to the total probability of coverage are weighted by the PDFs $\fr{.}{\los_\ess}$ and $\fr{.}{\nlos_\ess}$.

\subsection{Statistics of Actual Deployments}\label{dual_stats}
To assess how well our model matches with an actual two-operator deployment of comparable network densities and overlap, we first need to estimate these parameters from the actual deployment. Unfortunately, there is one realization of the point process we seek to model per market, so the first order (the densities) and second order statistics (the overlap) need to be estimated from this single realization. Hence, we conduct our statistical analysis on a single observation and assume that the underlying point process $\Phi$ is a PPP which is stationary and ergodic \cite{stoyan}. The key quantities to estimate given a bounded window $W$ are $\lambda$, $\lambda_1$, $\lambda_2$, and $\rho$. A general unbiased estimator of the total density $\lambda$ according to \cite{stoyan} is 
\begin{equation}\label{estimate_lambda}
\hat{\lambda} = \frac{\Phi(W)}{\vol{W}},
\end{equation}
and similar expression for $\lambda_1$ and $\lambda_2$ follow.

We also provide two ways to estimate the overlap coefficient given in \eqref{overlap_coef}, which involves a cross-moment of two point processes. Second-order moments of point processes and methods of their estimation are well studied in \cite{ripley76, ripley77, osher, jarvi} and the references therein, yet there is very little in the stochastic geometry literature on cross moments \cite{perkel, ripley76} and their estimation. Therefore, we devise estimating the overlap coefficient in one of two ways:
\paragraph{Indirectly through estimating $\lambda_{12}$} This uses the fact that the overlap coefficient $\rho$ is directly proportional to the density of the mother process $\Phi$ when the latter is Poisson, with the proportionality constant being $\lambda_{12}$. Hence, we compute $\hat{\lambda}_{12}$ as in ~\eqref{estimate_lambda}, and then set $\hat{\rho}=\frac{\hat{\lambda_{12}}}{\hat{\lambda}}$. 

\paragraph{Directly through a naive estimator} This extends the notion of sample covariance to point processes to estimate the overlap $\rho$ between two point processes. We first apply a uniform partition $\winofn$ of size $n$ to the observation window $W$. Finally, we compute $\hat{\rho}$ as:
\begin{align}
\begin{split}\label{empirical_overlap}
\rhohat_n &= \displaystyle \frac{\sum_{w\in\winofn} \Phi_1(w)\Phi_2(w) - \hat{\lambda}_1\hat{\lambda}_2|W|}{\hat{\lambda}|W|},\\
\rhohat &= \lim_{n\to\infty} \hat{\rho}_n.
\end{split}
\end{align}
Note that the expression of $\rhohat$ as given by \eqref{empirical_overlap} does not necessarily guarantee that $\lamwhat + \lamthat -\rhohat\lamhat = \lamhat$, which is the case when the actual deployment patterns are not Poisson. 

\subsection{System Comparison}
We consider two perspectives to study how coverage probability changes as a function of overlap between two mmWave cellular networks. For instance, increasing overlap does \emph{not} imply that individual densities need to increase as well. To make the comparison simpler, we assume that the two networks have equal BS densities.

\paragraph{Fixed individual network densities (FID)} Each operator chooses to share sites with the other operator while maintaining the density of its BSs. In a practical sense, this might occur when the lease between the operator and landlord/property owner is terminated or no longer renewed, forcing the operator to relocate its equipment to another site for continued coverage of the area it had been serving. A \emph{telecommunications lease} grants both the operator (lessee) and the landlord (lessor) the right to terminate the contract before the end of the lease term, or to opt out of renewing the lease. The operator can cancel the lease if it deems the property no longer technologically or economically suitable. The landlord can cancel the lease if the operator violates the terms of the lease by failing to obtain permits, governmental authorizations, and other approvals for utilizing the property, or failing to use the property as intended. Applying this to our model, consider, for example, two operators with fixed densities $\lambda_1 = \lambda_2 = \lambda_0$. To achieve an arbitrary overlap of $\rho$, each of the networks \emph{relocates} a number of BSs accounting for $\frac{\rho}{1+\rho}$ of its density into the same number of sites of the competing network. As a result, $\lambda_1'=\lambda_2'=\left(\frac{1-\rho}{1+\rho}\right)\lambda_0$ and $\lambda_{12}=\left(\frac{2\rho}{1+\rho}\right)\lambda_0$. We will refer to $\lambda_{12}/\lambda_0$ as the sharing ratio.

\paragraph{Fixed combined network density (FCD)} Each operator can share sites with the competing operator by means of \emph{expanding} into the sites owned by the latter; i.e., each operator retains the sites that it started out with. In a practical sense, this might occur as strategic action to extend an operator's reach in a market. For example, consider as above two operators with base densities $\lambda_1 = \lambda_2 = \lambda_0$. To achieve an overlap of $\rho$, each operator expands into $\rho$ of the competitors network. In this case, $\lambda_1'=\lambda_2'=(1-\rho)\lambda_0$ and $\lambda_{12}=2\rho\lambda_0$. Note that the highest density achieved by either operator is the sum of the individual starting densities of individual networks; i.e. $2\lambda_0$.

The main takeaway here is that increasing $\rho$ under $\stra$ increases the overlap but conserves the BS densities of each of the two networks. In contrast, increasing $\rho$ under $\strb$ increases the overlap \emph{and} the BS densities of the two networks. Additionally, increasing overlap decreases the overall density of the cell sites under $\stra$ but conserves the overall density under $\strb$.

Infrastructure sharing is, of course, contingent on the approval of regulatory bodies which weigh consumer gains and overall positive outcomes against harmful competition (i.e. competition that jeopardizes the competitors' revenues). One the one hand, passive sharing of sites and towers may ease the roll-out and expansion of an entrant by granting them access to key sites, but the entrant can grow in the long run into a strong competitor that threatens the market share of the incumbent. This encourages incumbents to share only a part of their infrastructure in such a scenario, with less sites being shared after an initial phase. On the other hand, a bigger operator could gain access into key sites owned by a smaller operator (for e.g., hilltop or downtown) and drive the latter out of the market. In both scenarios, either operator could opt for partial co-location, i.e. $\rho < 1$, to contain the competition. 
\section{Numerical Results}\label{num}
In this section, we present numerical results that validate our coverage analysis, evaluate the accuracy our model, and compare performance metrics across a range of overlap extents. We give all numerical results for the two-operator mmWave system.

For these results, we assume that the combined bandwidth of the mmWave systems operated by the networks is 200 MHz, and that the operating frequency is 28 GHz. The values of the rest of the parameters are similar to their counterparts in \cite{bai15,abi15}. For the power law path loss model, we consider $\clos=-60$ dB, which is the approximate close-in free space (Friis) path loss at a close-in reference distance of 1 m at 28 GHz. We assume that there is a fixed additional 10 dB power loss for NLOS links, i.e. $\cnlos=-70$ dB (see Table VI of \cite{rappaport15}). As for path loss exponents, we set $\alos=2$, and $\anlos=4$ according to \cite{rappaport15,sun16}. Additionally, we consider a transmit power of $26$ dBm,  and a standard thermal noise power spectral density of $-174$ dBm/Hz with a noise figure of $10$ dB. As for the parameters of the sectored antenna model, we assume that the BS is equipped with an $8\times 8$ planar antenna array which has a corresponding main lobe gain of $\sfgmax=18$ dB. As for the side lobe gain, we choose $\sfgmin=-2$ dB as value that approximates the magnitude of the radiation pattern of the antenna beyond the main lobe, i.e. side lobes, nulls, and everything in between. Additionally, we select a half beam-width of $\theta_b=10^{\circ}$. As for the densities of the two networks, we consider a reference density of $\lambda_0 = 30$ per km$^2$ which is equivalent to a cell radius of $103$ m. Each network has an associated \emph{active} user density of $200$ per km$^2$. For the mmWave exponential blockage model, we consider $\beta = 0.007$ corresponding to an average LOS region of 144 m.

\subsection{Validation of SINR Coverage Analysis}
\begin{figure}[h]
\centering
\includegraphics[width=0.49\textwidth]{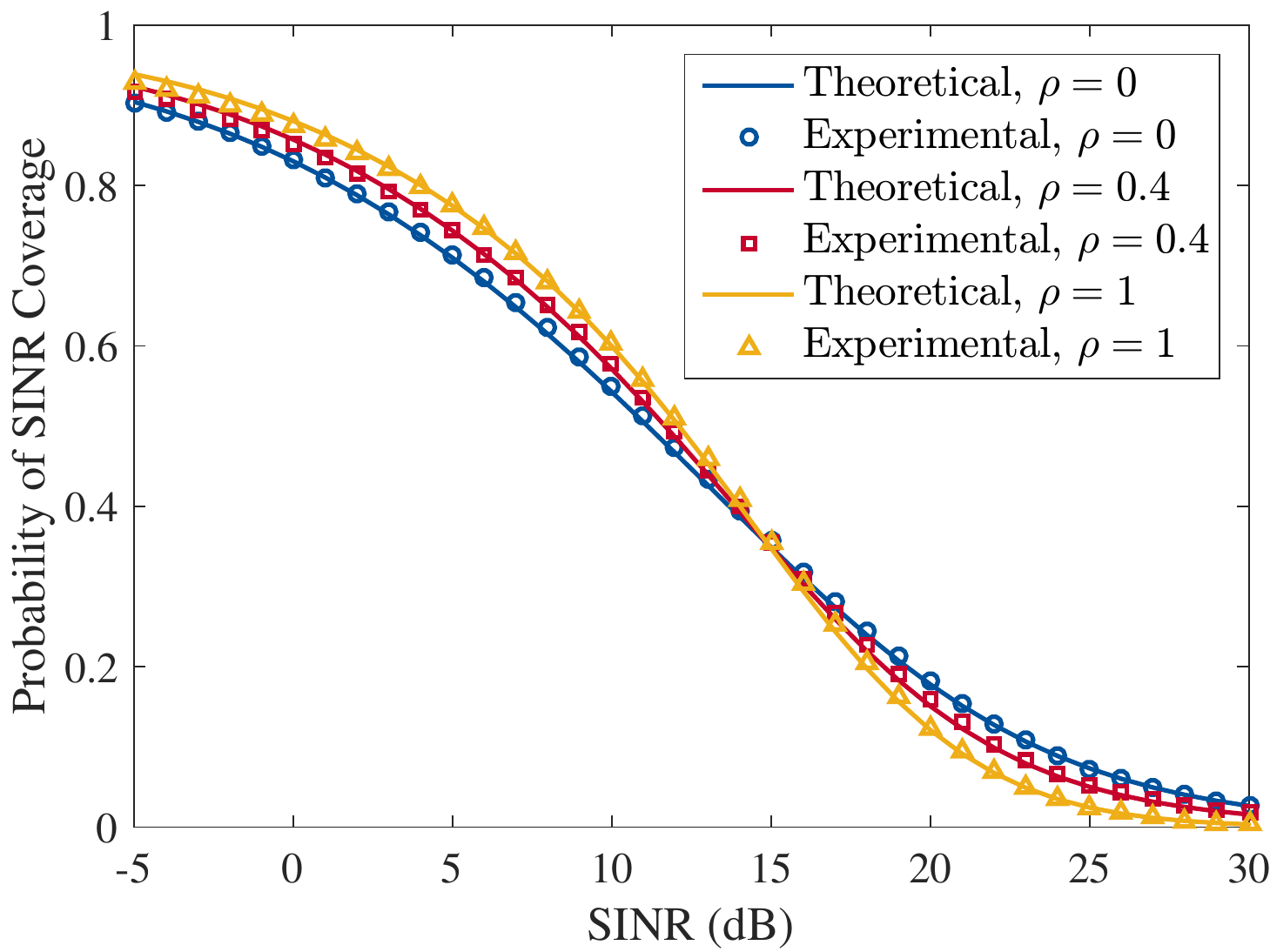}
\caption{Probability of SINR coverage vs. SINR threshold (dB)  for different sharing schemes with individual network densities fixed at $\lambda_0$. Solid curves correspond to analytical results and marked curves correspond to simulation results.}
\label{fig_sinr_validate}
\end{figure}

We validate the analytical expressions for the probability of SINR coverage that were obtained in Section \ref{dual_model} for a two-operator mmWave system with closed access and full spectrum sharing. We numerically evaluate the probability of coverage expression in \eqref{eq_pcov_dual} for a range of SINR thresholds, and we plot this against the empirical probability obtained through Monte Carlo simulation; the results are shown in Fig.~\ref{fig_sinr_validate}. We consider $\stra$ with different overlap coefficient values, $\rho=0$ or no infrastructure sharing, $\rho=0.4$ or 57\% sharing, and $\rho=1$ or full sharing (by letting $\rho$ grow large, the two networks share more sites in common but their individual densities remains constant throughout). The first thing we observe, for all considered degrees of sharing, is that plots obtained through simulation match the ones obtained through analysis; which further validates the correctness of our analysis. Moreover, we observe that increasing the overlap decreases the coverage probability at higher SINR thresholds, yet increases the coverage probability at lower SINR thresholds. The reason is that, in full sharing, there are no interfering BSs closer to the user than the associated BS, which has a positive impact on coverage at low SINR thresholds. As for high SINR, the anticipated signal is received at a much higher power than that of the interfering signals combined, but this is not the case in full sharing; the associated BS is co-located with another BS, which adds yet a source of interference that is just as powerful as the intended source.

\subsection{Comparison of Estimators for the overlap $\rho$}
\begin{figure*}
\centering
\includegraphics{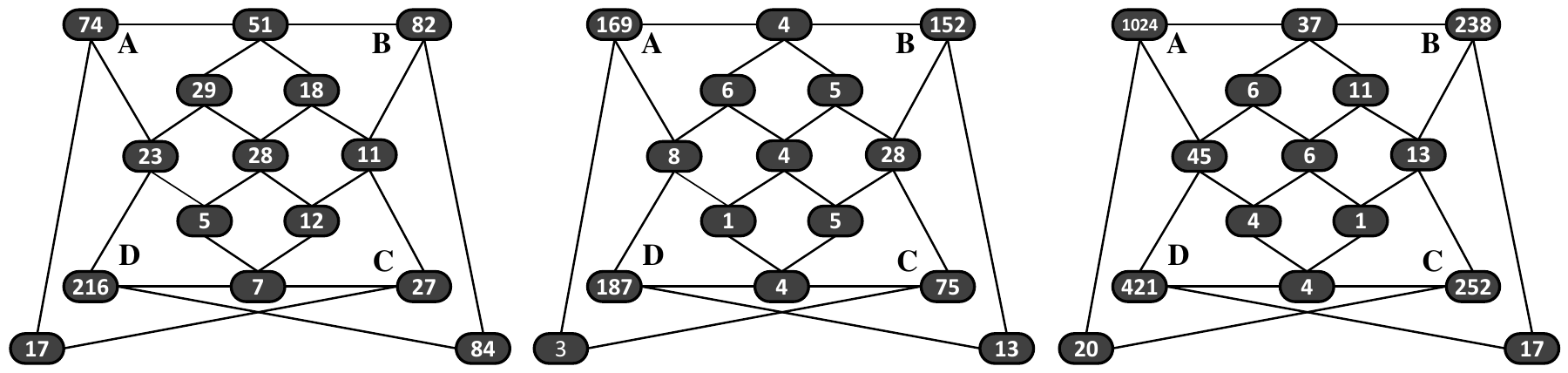}
\caption{Hasse Diagrams showing the number of macro towers and other structures that are shared by the four largest US operators in three select CMAs, \cmai/, \cmaii/, and \cmaiii/. Nodes in these diagrams are marked by elements in the power set of $\set{\text{A, B, C, D}}$. Each of these elements/subsets denote the identity of operators whose BSs are co-located with one another, and the number inside the corresponding node refers to the number of these shared structures. As an example, the corner nodes,  marked with letters corresponding to the sets $\set{\text{A}}$, $\set{\text{B}}$, $\set{\text{C}}$ and $\set{\text{D}}$, contain the number of towers occupied by BS antennas of \emph{one and only one} operator. Moving inwards, edges represent set membership; equivalently, moving outwards, edges represent set inclusion. As another example, in the leftmost diagram, $|\set{\text{A}}|=74$, $|\set{\text{B}}|=216$, $|\set{\text{C}}|=27$, $|\set{\text{A},\text{D}}|=23$, $|\set{\text{C},\text{D}}|=7$, and finally $|\set{\text{A},\text{D},\text{C}}|=5$; where the final term represents the towers with BSs of A, C, D \emph{but not} B.}
\label{hasse_diagram}
\end{figure*}

We compare the direct and indirect estimators for $\rho$ proposed in \ref{dual_stats}. Since there are no current mmWave BS deployments, we have obtained the coordinates of current BS locations of legacy networks (2G to 4G) for the four largest US operators in three major CMAs of different geographic and demographic characteristics. The Hasse diagram in Fig.~\ref{hasse_diagram} shows the number of macro towers and other structures (rooftops, stealth, and DAS nodes) that house BS antennas of the different operators in every market. While there appears to be little infrastructure sharing in \cmaiii/, \cmai/ and \cmaii/ display strong instances of sharing.

To evaluate the accuracy of our two-operator PPP model, we compare the probability of SINR coverage between the PPP model and actual two-operator networks in different markets. Since mmWave systems have not been deployed yet, we ``down-scale'' the abscissas and ordinates of the BS locations so that the individual network densities compare to that of a typical mmWave network. We set that to 60 BSs/km$^2$. This operation is known as \textit{pressing}, and is essentially an affine transformation in the plane which maintains the Poisson property of the original, full-scale point process. Note that pressing a network of two operators maintains their overlap.

\begin{figure}[h]
\centering
\subfloat[\cmai/]{
	\includegraphics[width=0.48\textwidth]{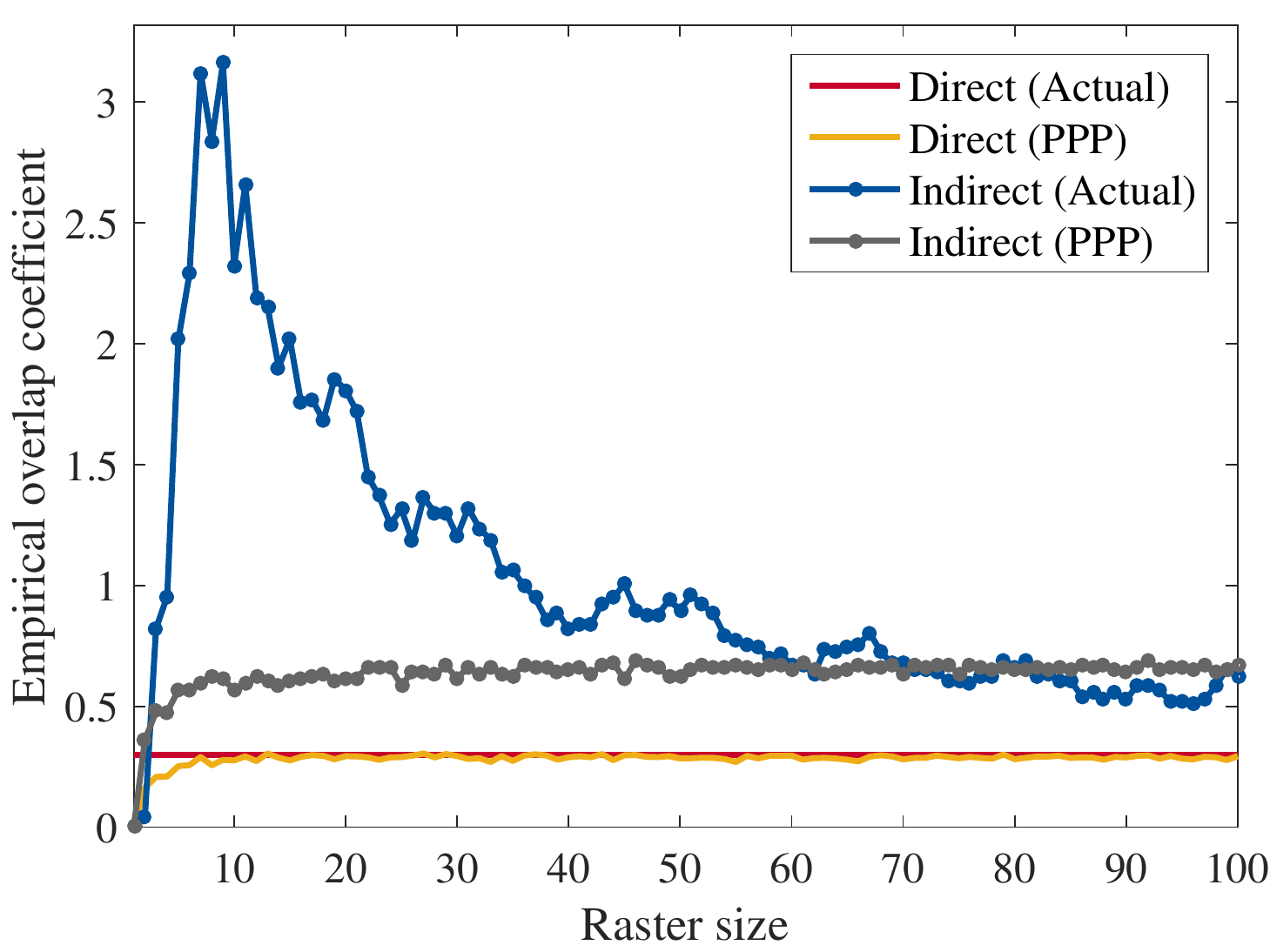}
}
\hfill
\subfloat[\cmaii/]{
	\includegraphics[width=0.48\textwidth]{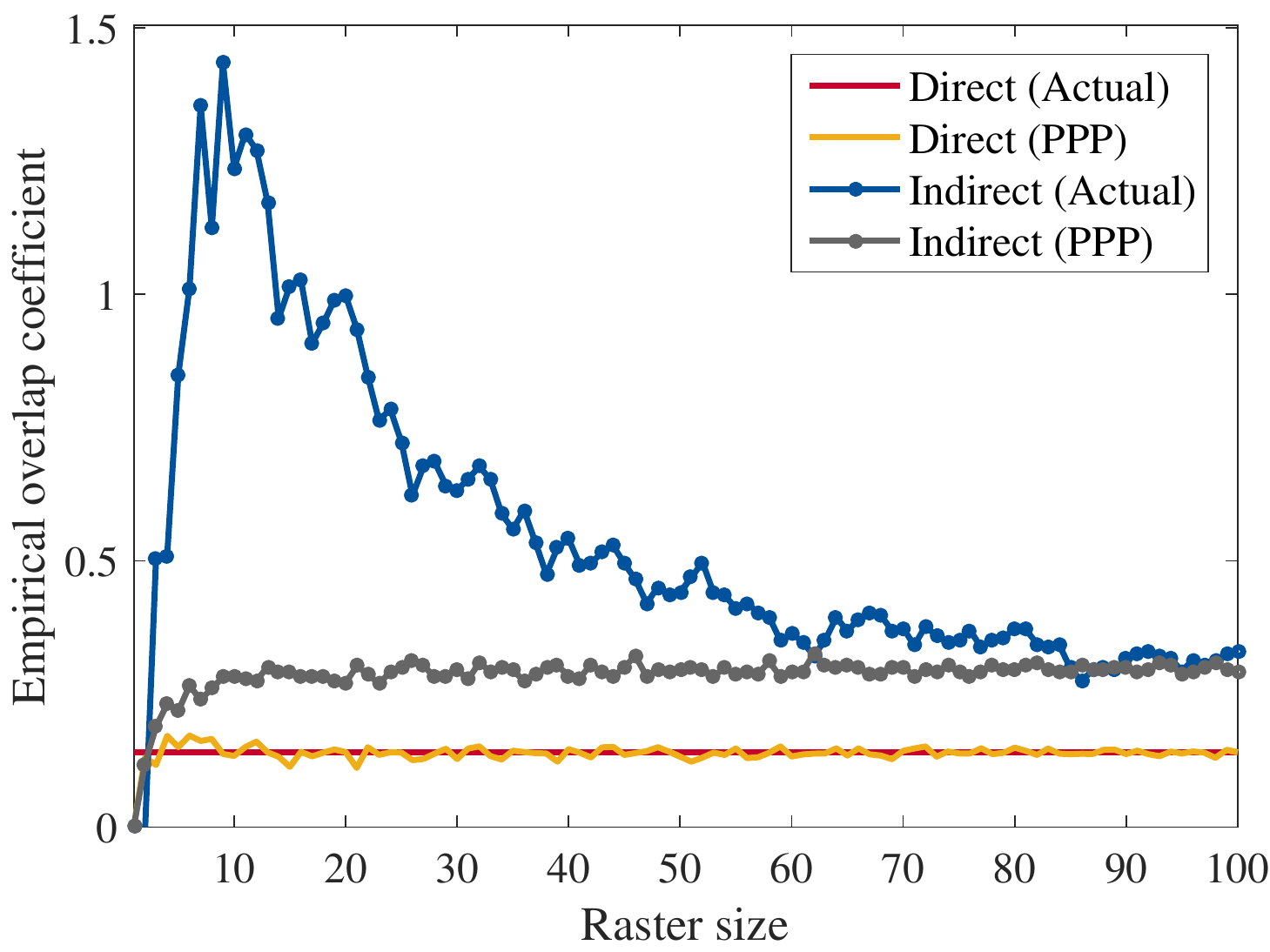}
}
\caption{Comparison of direct and indirect estimators for the empirical overlap coefficient, $\rho$, as computed from actual data, and from PPP realizations. For an actual deployment, the direct estimator is filtered with a simple moving average to highlight its general trajectory as a function of the size of the raster that divides the observation window.}
\label{fig_empirical_overlap}
\end{figure}

For each market, we first estimate the densities and overlap using \eqref{estimate_lambda} and \eqref{empirical_overlap}. Then, we generate a number of realizations of our two-operator PPP model with the obtained estimates. Finally we \emph{re-estimate} the densities and overlap of these realizations. The reason for this procedure is two-fold. First, mismatching estimates for $\rho$ suggest that BS locations of actual deployments are not really PPP realizations. Second, we can verify that \eqref{empirical_overlap} is in fact an estimator for $\rho$ at least for a PPP. We plot our findings in Figure~\ref{fig_empirical_overlap}. We filter the direct estimate of $\rho$--which is a quantity that evolves with the number of bins--with a simple moving average filter to highlight the general trend it follows. The first observation we make is that the two estimates are different in both datasets: the two estimates are off by 0.2 in the first dataset and by 0.1 in the second. This suggests that the BS locations are not quite PPP. The second observation is that difference of the two estimates for \cmaii/ is less than the error for \cmai/. This can be explained by the fact that \cmaii/ is a dense, DAS-node-dominated network that extends over 1\% the area covered by \cmai/ which is a macro-tower-dominated network; and DAS node locations are more random than those of cell towers.

\subsection{Validation of Model with Actual BS Deployments}

\begin{figure}[h]
\centering
\subfloat[\cmai/: $70$ km $\times$ $70$ km window of an actual macro site deployment shared by \opa/ and \opb/.]{
\includegraphics[width=0.45\textwidth]{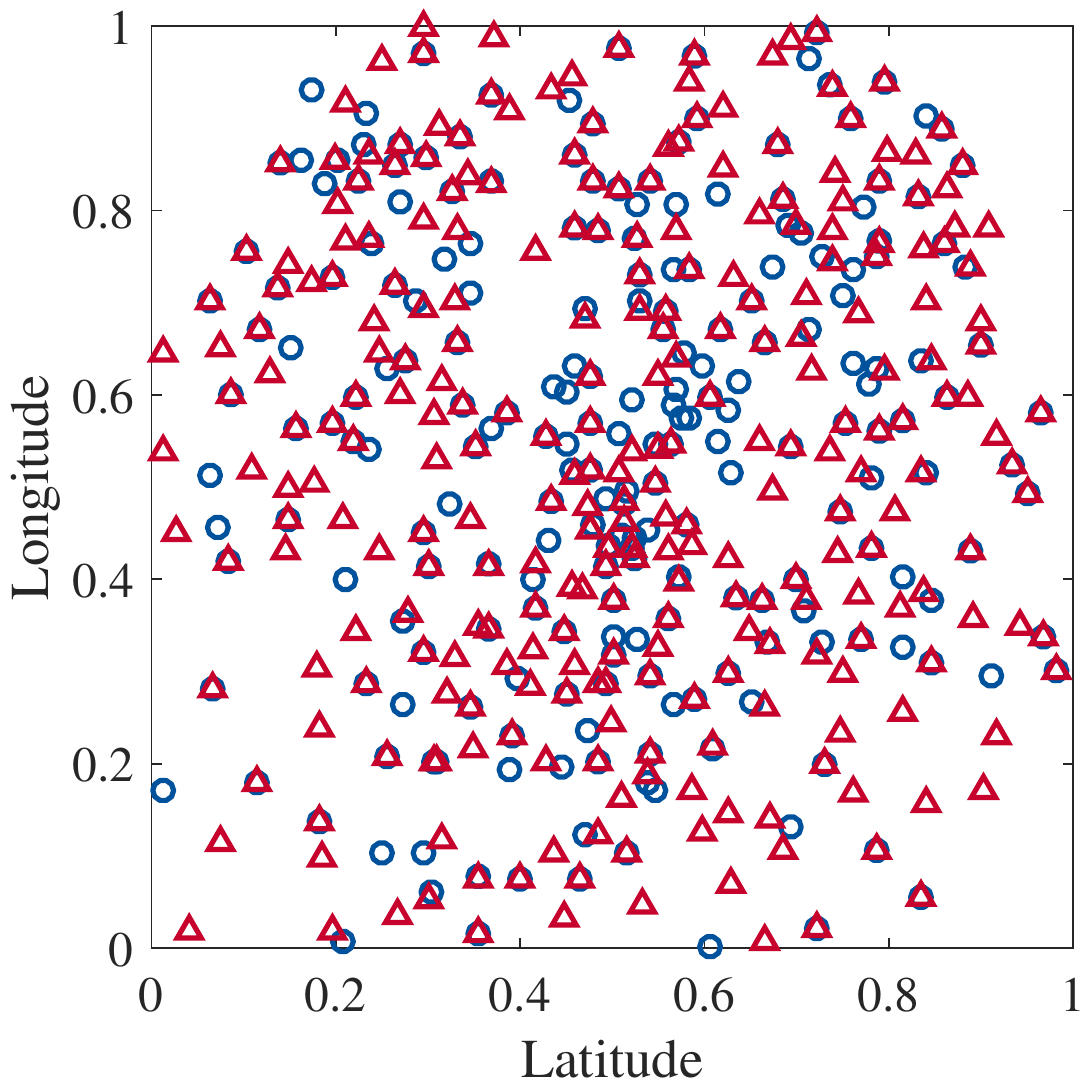}
}
\hfill
\subfloat[\cmaii/: $20$ km $\times$ $20$ km window of an actual DAS-node-dominant network shared by \opb/ and \opc/.]{
\includegraphics[width=0.35\textwidth]{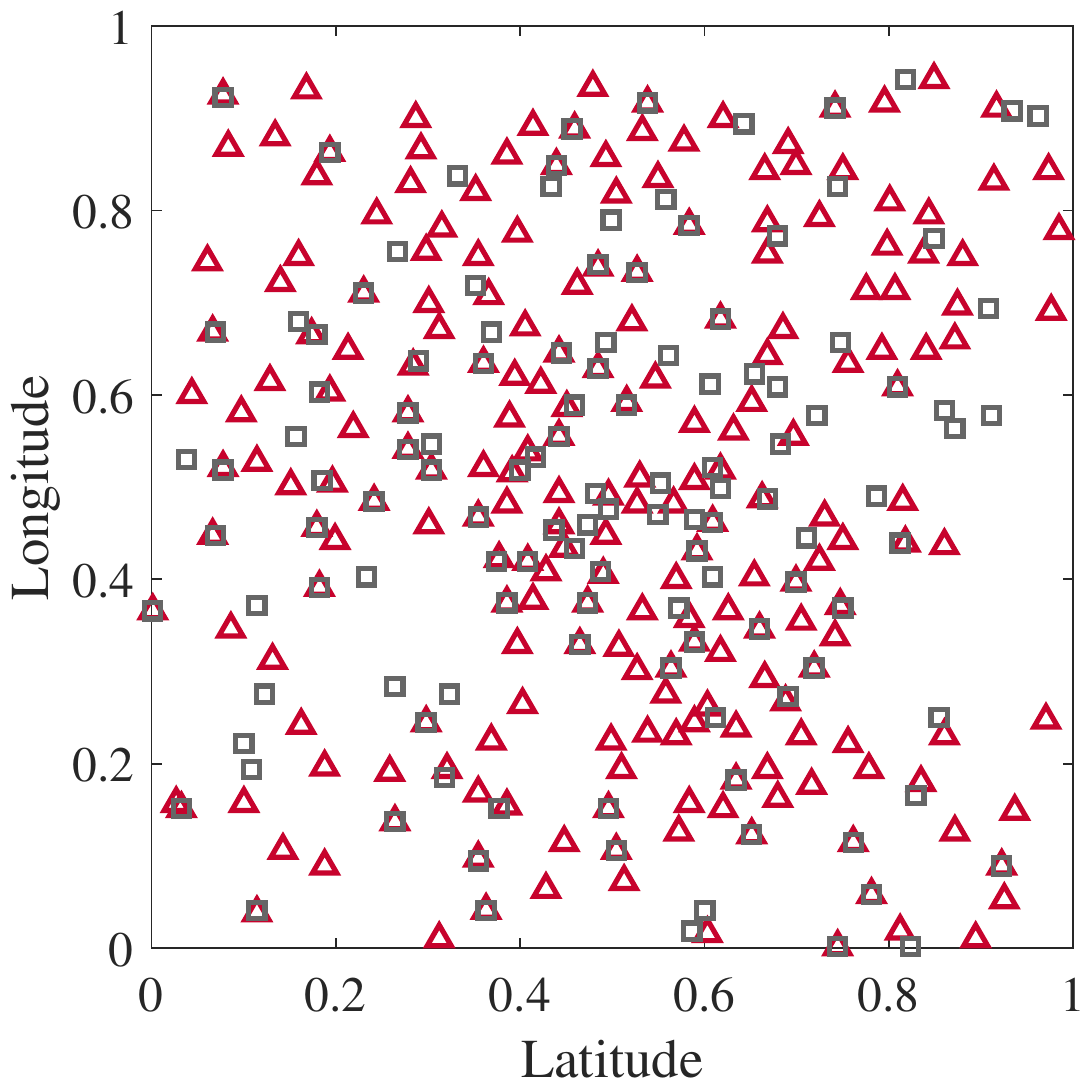}
}
\caption{Actual deployment of three major US operators in two major markets. \opa/ \protect\mycircle{clr_steel_blue}, \opb/ \protect\mytriangle{clr_steel_red} and \opc/ \protect\mysquare{clr_my_dgray}\;. \cmai/ extends over 100$\times$ the area of \cmaii/. Corner vacancies in the first figure correspond to outskirts of the market. Co-located BSs are represented by overlapping shapes.}
\label{actual}
\end{figure}

\begin{figure}[h]
\centering
\subfloat[\cmai/]{
\includegraphics[width=0.48\textwidth]{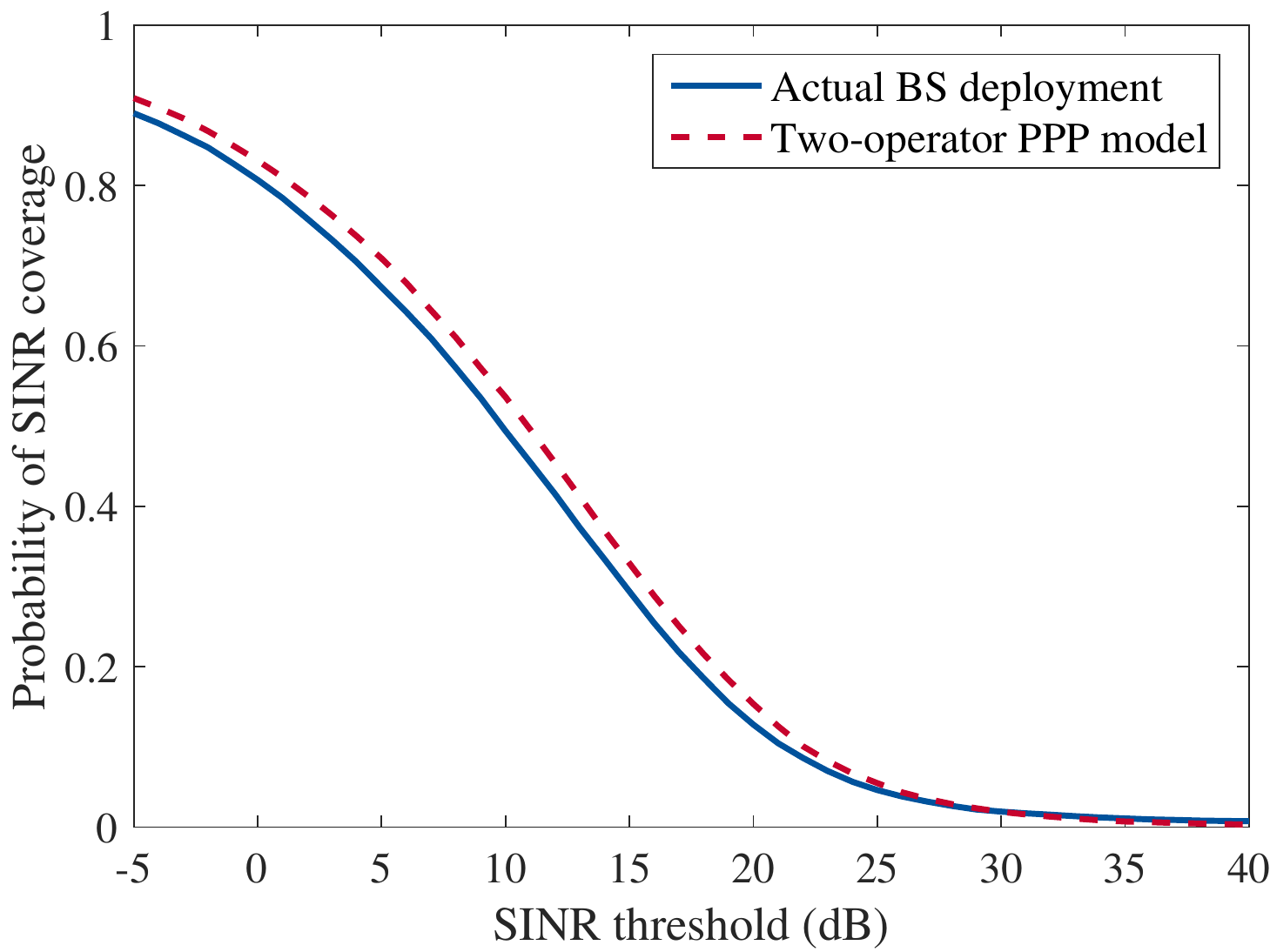}
}
\hfill
\subfloat[\cmaii/]{
\includegraphics[width=0.48\textwidth]{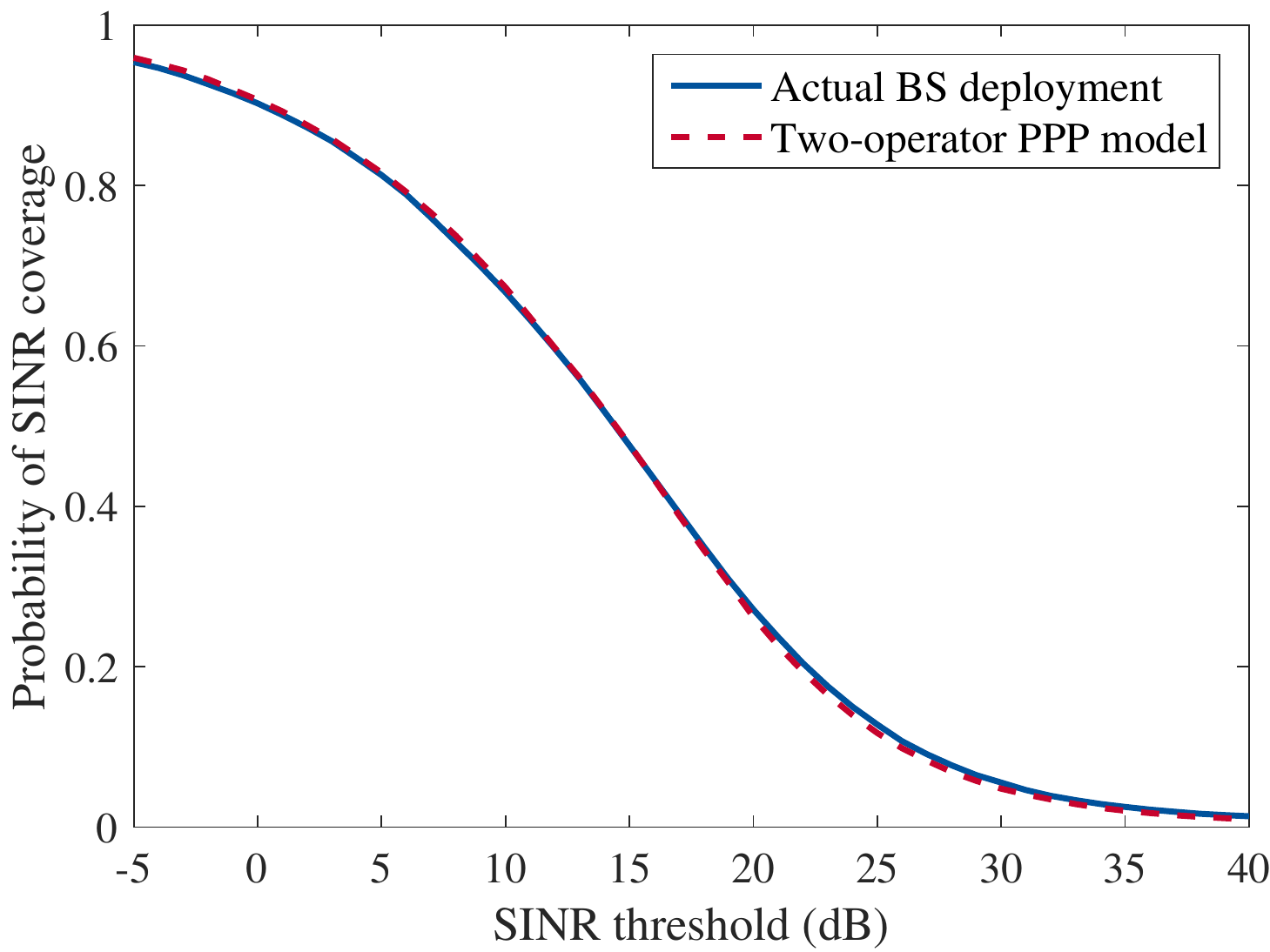}
}
\caption{Probability of SINR coverage vs. SINR threshold (dB). Probabilities are obtained for the two-operator PPP model with individual network densities of 60 BSs/km$^2$ and actual two-operator networks with matching densities. The networks in \cmai/ are macro-cellular, while those in \cmaii/ are predominantly DAS-node based. The two figures suggest that the PPP model would is very accurate denser network of randomly-positioned BSs.}
\label{actual_ppp}
\end{figure}

To assess how accurately our model reflects the performance of an actual two-operator system, we compare the probability of coverage of the two systems while constraining the two networks to have the same individual densities and overlap. We do this experiment for two sets of actual deployments.

We first estimate $\lambda_1$, $\lambda_2$ and $\lambda_{12}$ as described in Section~\ref{dual_stats}. Figure~\ref{actual} shows the views of base station deployments in \cmai/ and \cmaii/. The figure on the left shows base stations of networks A and B, and the one on the right shows those of \opb/ and \opc/.  Figure~\ref{actual_ppp} shows the SINR probability of coverage attained by a typical user connecting to \opa/ (left) and \opc/ (right). Note that the overlap coefficent estimates used in evaluating the SINR coverage probability formula were obtained indirectly. We can see that the SINR coverage plots for our two-operator PPP model match reasonably well with those for the actual deployments. Even though our model produces an almost identical coverage curve for \cmaii/, it produces a coverage probability curve that slightly deviates from the actual deployment at a wide range of SINR.

\subsection{Comparison of Sharing Schemes}
\begin{figure*}[h]
\centering
\subfloat[Rayleigh fading]{
	\includegraphics[width=0.5\textwidth]{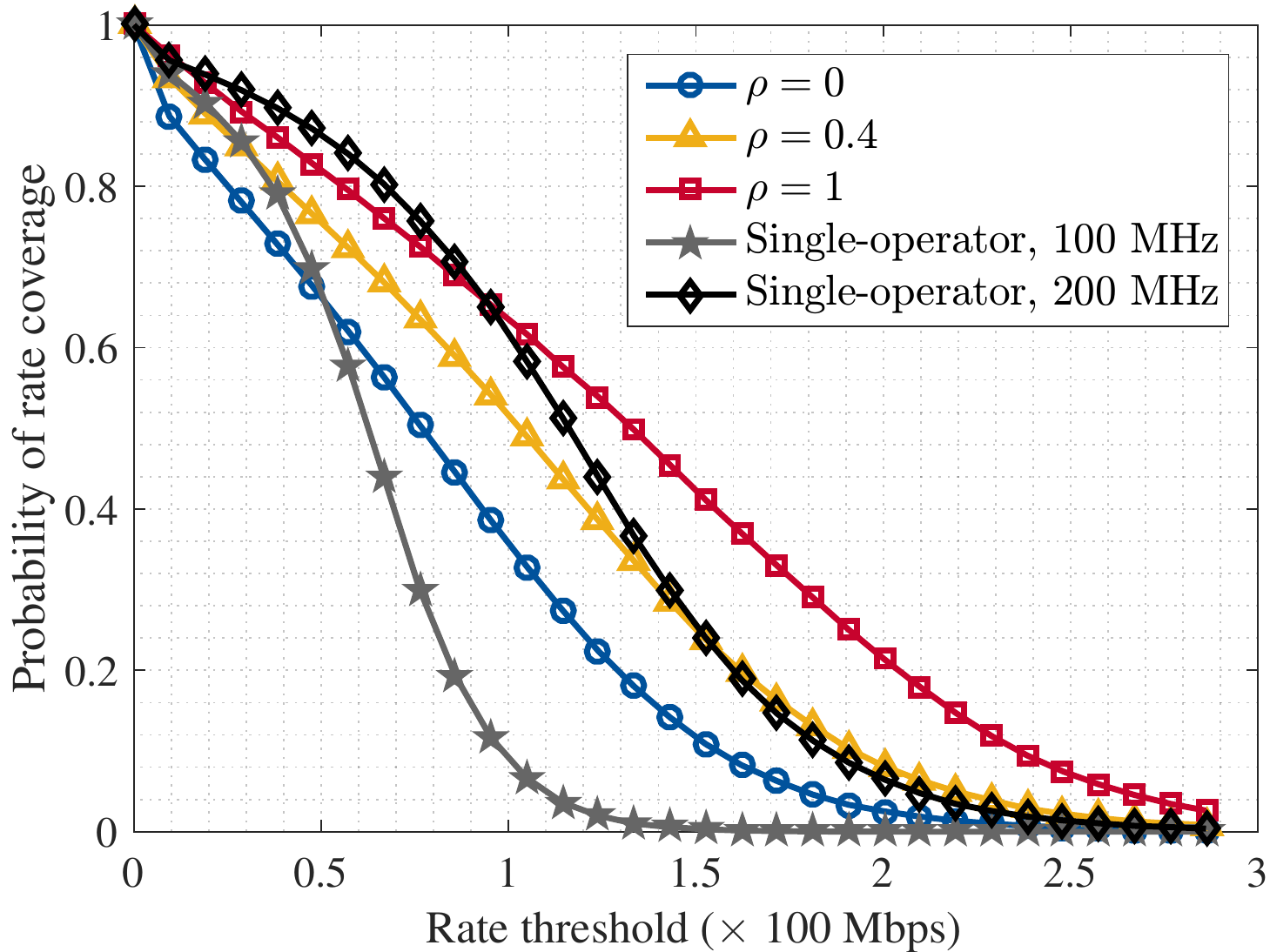}
}
\subfloat[Nakagami fading and log-normal shadowing]{
	\includegraphics[width=0.5\textwidth]{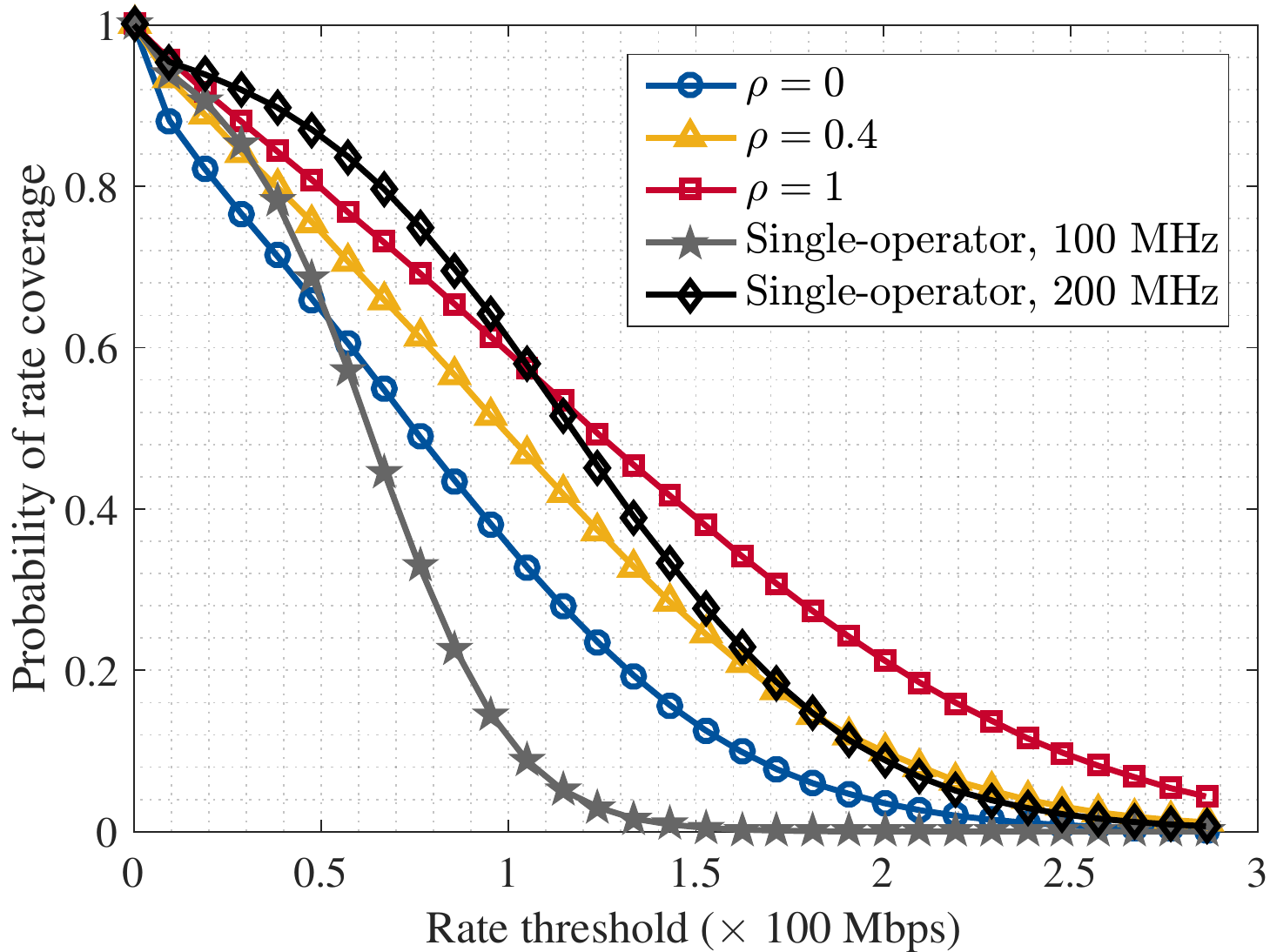}
}
\caption{Probability of rate coverage vs. downlink rate threshold ($\times 100$ Mbps) for shared-networks with $\rho=$ 0, 0.4, and 1 under FCD sharing strategy, a single-operator network with 100 MHz of bandwidth, and a single-operator network with 200 MHz of bandwidth. Combined network density is fixed at $\lambda_0$ = 60 BSs per km$^{2}$. Coverage probabilities were computed experimentally with Rayleigh fading (Left), and Nakagami fading with parameters $m=2,3$ for LOS and NLOS, and log-normal shadowing with power $\sigma_{\text{dB}}=5.2,7.6$ for LOS and NLOS (Right). The trends in the two figures are identical which justifies using Rayleigh fading assumption in our analysis.}
\label{rate_fcd}
\end{figure*}

\begin{figure*}
\centering
\includegraphics[width=0.5\textwidth]{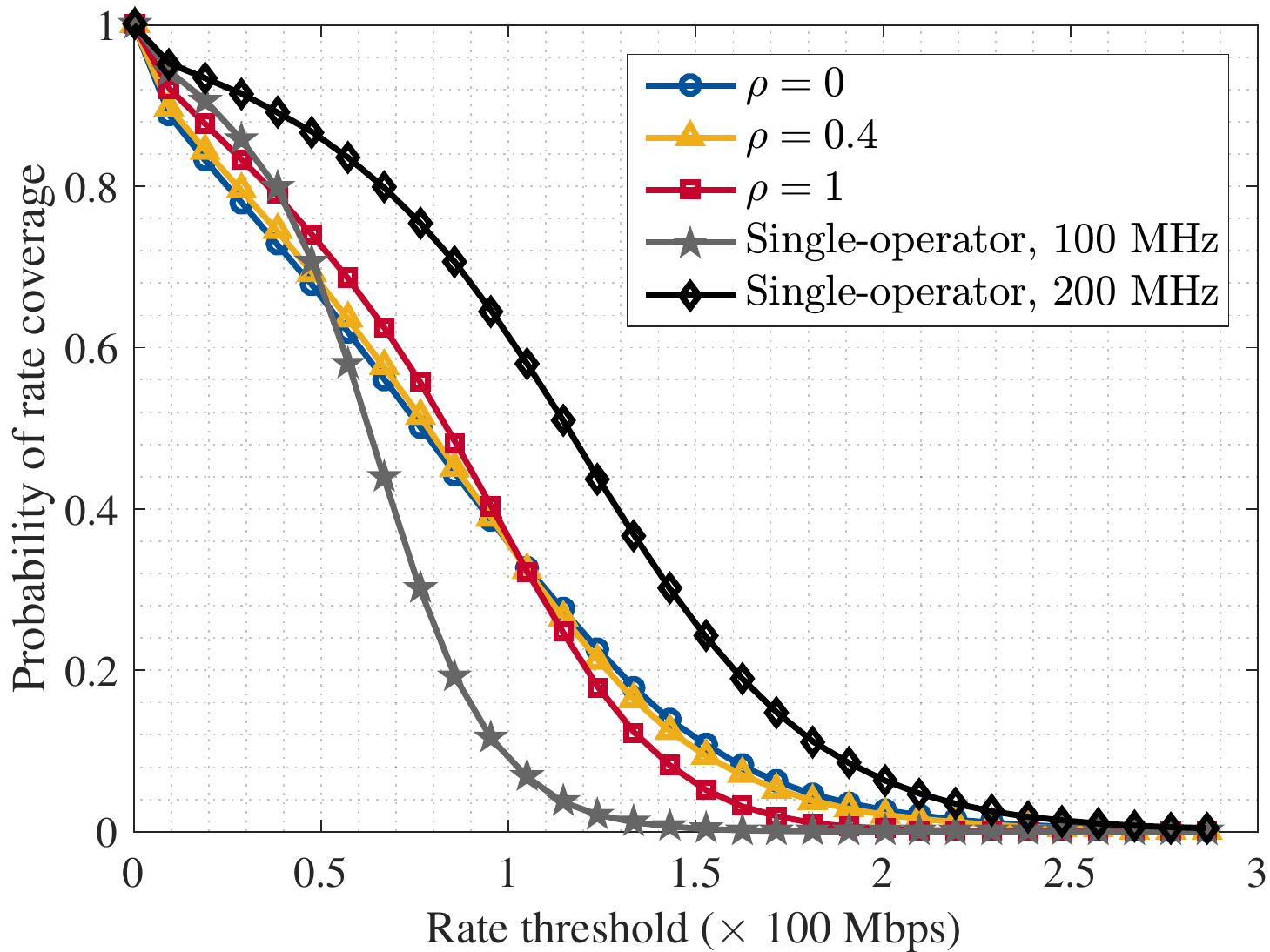}
\caption{Probability of rate coverage vs. downlink rate threshold ($\times 100$ Mbps) for different values of $\rho$ under FID with individual network densities fixed at $\lambda_0$ = 30 per km$^2$.}
\label{fig_rate_fid}
\end{figure*}

\begin{figure}[h!] 
\centering
\includegraphics[width=0.5\textwidth]{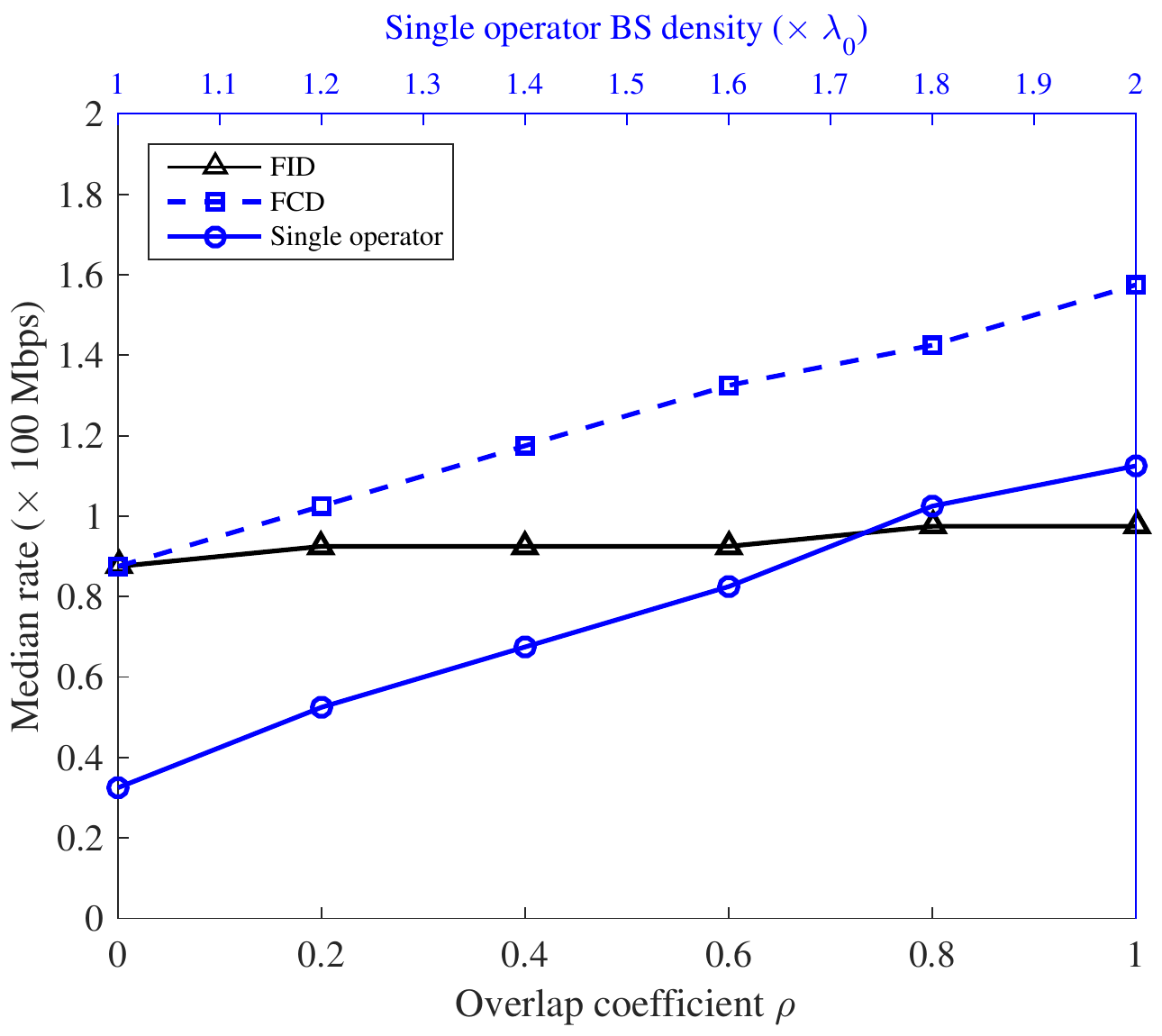}
\caption{Comparison of median rate: FID strategy, FCD strategy, and single operator system. For a single operator, median rate is a function of the network's density. For FID, individual network densities is fixed at 30 per km$^2$; hence, median rate is a function of the overlap coefficient. For FCD, individual network densities grows with overlap. Hence median rate varies with both overlap and individual densities.}
\label{fig_median_rate}
\end{figure}

We compare the probability of rate coverage between three shared networks with different overlap as well as two single-operator networks with different bandwidth sizes under $\stra$ and $\strb$. A shared network is a network that shares infrastructure and spectrum with another network, while a single-operator network is one that shares neither infrastructure nor spectrum. For shared networks, we consider $\rho = $ 0, 0.4, and 1 and a combined fixed density of $2\lambda_0$. Whereas for single-operator networks, we consider 100 MHz and 200 MHz of available bandwidth and a fixed density $\lambda_0$.

Figure~\ref{rate_fcd} compares the probability of rate coverage between the different networks under $\strb$. We first compare the performance of shared networks to a single-operator network with 100 MHz of bandwidth. We observe that fixing the total density to a base value $\lambda_0$ and increasing the overlap increases the probability of coverage for all rate thresholds. This is due to the fact that as the overlap increases, the densities of the operators' networks increase, and as a network's density increases, less number of users load a BS.  At low rate thresholds, a single-operator network with 100 MHz of available bandwidth outperforms a fully-separated network (the case when $\rho=0$). Even though the single-operator network has half the bandwidth, the interference that a fully-separated network experiences from the other network with whom it shares spectrum can be so significant that it deteriorates performance at low thresholds. The single operator network competes with the partially-shared network (the case when $\rho=0.4$), and there are three factors that explain their similar performance. First, the density of the partially-shared network is higher by virtue of sharing some of the sites of its peer network, which reduces the load on its BSs and promotes higher per-user rates. Second, the partially-shared network has double the bandwidth. Third, despite that the single-operator network has a higher load on its BSs and less spectrum resources, there are no interfering BSs that are closer to the user than their associated BS, which leads to a higher probability of coverage at low SINR thresholds. However, the single-operator network underperforms fully-shared networks (the case when $\rho=1$). Contrary to before, there are no interfering BSs that are closer to the user than their associated BS in both of these networks. Fully-shared networks provide a higher \emph{rate} coverage probability simply because they have twice the bandwidth that is available to the single-operator network and around half the load. Additionally, Figure~\ref{rate_fcd} shows that if two operators share only spectrum, their networks will provide up to 500\% increase in rate coverage which is achieved at a threshold of 100 Mbps. If each network further expands into 40\% of the competitors' network, i.e. an overlap coefficient of $0.4$, their networks will provide up to 800\% increase in rate coverage which is achieved at a threshold of 100 Mbps. 

We now compare the performance of shared networks to a single-operator network with 200 MHz of bandwidth. We observe that the single-operator network with 200 MHz of bandwidth outperforms all other networks \emph{at low rate thresholds.} In particular, it outperforms a fully-shared network because it is not subject to internetwork interference. It also outperforms partially-shared and fully-separated networks because of the absence of interfering BSs that are closer to the user than the associated BS. Nevertheless, the reduced load on the BSs of the shared networks appears to overcome the effect of low SINR due to near as well as co-located interferers which increases the rate coverage probabilities of these networks at high rate thresholds.

Finally, we highlight that the trends in the probability curves are identical regardless of the assumption of small-scale and large-scale fading.

Figure~\ref{fig_rate_fid}  compares the probability of coverage between the same shared and single-operator networks under the $\stra$ perspective, and two important observations can be made. First of all, the probability of rate coverage does not change appreciably with varying overlap. This can be explained by the fact that each of the networks maintain its individual density and, thus, the load on its BSs. Of course, higher coverage at low rate thresholds for deployments with more prominent sharing and the opposite trend at high thresholds follows the same reasoning we made before in relation to Figure~\ref{fig_sinr_validate}. Second, the single-operator network with 100 MHz of bandwidth outperforms all shared networks for low rate thresholds, which suggests that it is better suited to provide services to devices that communicate through very low rates, such us IoT devices that wake up infrequently to send status updates or small measurements. Last but not least, we observe that the single-operator network with 200 MHz outperforms all other networks. This is explained by the combined benefit of having maximum bandwidth and having no inter-network interference, as is the case for all shared networks, particularly near interferers, as is the case for the partially-shared and fully-separated networks.

Figure~\ref{fig_median_rate} shows the trajectory of the median rate of the two sharing strategy and the reference strategy with the increase of either of the overlap coefficient (FID), or individual operator density (single operator), or both (FCD). For $\stra$, unlike $\strb$, increasing the overlap between the two operators does not increase individual densities. For the single operator case, the operator shares neither spectrum nor infrastructure with other competitors. We observe that $\strb$ gives a higher rate than the two strategies. In particular, the median rate an operator achieves under $\strb$ is almost double the median rate achievable by a single operator at any individual density. Even though the interference in $\strb$ is higher due to the presence of transmitting base stations in equal density for the second operator, the used spectrum is double. Moreover, sharing gives an almost steady median rate for $\stra$. The reason is that the number of interfering BSs of the opposite network remains the same with varying the overlap coefficient. The only difference that changing overlap makes is whether the interferers of the opposite network are co-located with those of the home network.

\section{Conclusion}\label{conc}
In this paper, we proposed a novel mathematical framework for modeling BS locations of a multi-operator cellular mmWave system, and provided analytical expressions for the SINR coverage probability as a performance metric for an arbitrary network. For a tractable evaluation of system performance, we narrowed the scope to a two-operator scenario. For the two-oprator scenario, we provided a method to fit actual deployments with our model and estimate its necessary parameters, and validated the model with realistic deployments of cellular systems. Additionally, we suggested different schemes that describe increased infrastructure sharing: FID and FCD. We observed that the median rate in FID does not change appreciably as the overlap between the networks varies, unlike in FCD that witnesses a steadily increasing median rate in parallel to increasing overlap (or equivalently, increasing the density of the individual networks). We saw that the FCD strategy outperforms the single operator system for the same BS density, which is possible due to the availability of double the spectrum resources. We also saw that varying the extent of co-location  hardly alters the probability of coverage under FID. Since infrastructure sharing additionally offers economic incentives, it might be preferred to the case of no sharing. Finally, we observed that, for both FID and FCD, the single operator system with half the total bandwidth excels in the low rate threshold regime. That is, it is able to provide low data rates to more users than some of the two-operator systems under FID or FCD. 


\appendices
\section{Proof of Proposition~\ref{prop_interference_laplace}}\label{app_a}The user connects to $\netw$ through $\los_\tee$. The Laplace transform of $\ilos$ and $\inlos$, the interference from the LOS and NLOS blocks, is given as follows:

\begin{align*}
\ilos &= \sum_{\substack{\ess\in\peeof{\op}\\ 1\notin\ess}} \sum_{\substack{X_i \in L_\ess}} \sum_{l\in\ess} \clos H_{i,l} G_{i,l} ||X_i||^{-\alos} \\
		&+ \sum_{\substack{\essp\in\peeof{\op}\\ 1\in\essp}} \sum_{\substack{X_j \in L_{\ess'}}} \sum_{m\in\essp} \clos H_{j,m} G_{j,m} ||X_j||^{-\alos} \indic{X_j\in\ballc{r}} \\
		&+ \sum_{n\in \tee} \clos H_{0,n} G_{0,n} r^{-\alos}. \numberthis
		\label{eq_multi_laplos}
\end{align*}
\begin{align*}
\inlos &= \sum_{\substack{\ess\in\peeof{\op}\\ 1\notin\ess}} \sum_{\substack{X_i \in L_\ess}} \sum_{l\in\ess} \cnlos H_{i,l} G_{i,l} ||X_i||^{-\anlos} \\
		&+ \sum_{\substack{\ess'\in\peeof{\op}\\ 1\in\essp}} \sum_{\substack{X_j \in L_{\ess'}}} \sum_{m\in\essp} \cnlos H_{j,m} G_{j,m} ||X_j||^{-\anlos}\\
		&\hspace{3.2cm}\cdot\indic{X_j\in\ballc{D_L(r)}}. \numberthis
		\label{eq_multi_lapnlos}
\end{align*}

We derive the expression for $\lap{I_L}$ ($\lap{I_N}$ follows similarly). First, let $F_{\mathbf{u}}=H_{\mathbf{u}}G_{\mathbf{u}}$, for any index $\mathbf{u}$. Since the point processes representing the different brackets are independent by construction, the Laplace transform $\lap{I_L}(s)$ of $I_L$ is 
\begin{flalign*}
&\prod_{n\in \tee}\expect{\expp{-s\clos H_{0,n} G_{0,n} r^{-\alos} }}\\
\cdot &\prod_{\substack{\ess\in\peeof{\op}\\ 1\notin\ess}} \expect{\expp{-s\sum_{\substack{X_i \in L_\ess\\ l\in\ess}} \clos F_{i,l} ||X_i||^{-\alos} }} \\
\cdot &\prod_{\substack{\essp\in\peeof{\op}\\ 1\in\essp}}\hspace{-0.2cm} \expect{\expp{-s\sum_{\substack{X_j \in L_\essp\\ m\in\essp}}  \clos F_{j,m} ||X_j||^{-\alos} \indic{X_j\in\ballc{r}} }}.
\end{flalign*}
Taking the Laplace transform of PPPs with respect to the function $\clos F_{i,l}||X_i||^{-\anlos}$, and noting that independence of the fading and directionality gain RVs, and letting $q(\diff t)=2\pi t\plos(t)\diff t$, $\lap{I_L}(s)$ becomes equal to
\begin{align*}
&\prod_{n\in \tee}\expectw{G_n}{\lapof{H_n|G_n}{s\clos G_n r^{-\alos}}} \\
\cdot &\prod_{\substack{\ess\in\peeof{\op}\\ 1\notin\ess}} \expp{-\lambda_{\ess}\int\limits_{t\geq 0}\left(1-\Exp\prod_{l\in\ess}e^{-st^{-\alos}\clos F_l}\right)q(\diff t)} \\
\cdot &\prod_{\substack{\essp\in\peeof{\op}\\ 1\in\ess}}\hspace{-0.2cm} \expp{-\lambda_{\essp}\int\limits_{t\geq r}\left(1-\Exp\prod_{m\in\ess}e^{-st^{-\alos}\clos F_m}\right)q(\diff t)}.
\end{align*}
Moreover, since the independent marks $\set{H_k}$ and $\set{G_k}$ are assumed to be identically distributed, the Laplace transform simplifies as
\begin{align*}
\lap{I_L}(s) &=  u_L(s,r) ^{|\tee|-1} \\
\cdot &\prod_{\ess:\,1\notin\ess}   \exp\left(-\lambda_\ess \int\displaylimits_{t\geq 0} \left(1-u_L(s,t)^{|\ess|}\right) q(t)\diff t \right) \\
\cdot &\prod_{\essp:\,1\in\essp} \exp\left(-\lambda_\essp \int\displaylimits_{t\geq r} \left(1-u_L(s,t)^{|\essp|}\right) q(t)\diff t  \right).
\end{align*}
Finally, since the LOS and NLOS blocks are independent, we have that $\lapof{\ilos+\inlos}{s}=\lapof{\ilos}{s}\cdot\lapof{\inlos}{s}$.

\section{Proof of Proposition~\ref{dual_construction}}\label{app_c}According to the thinning theorem for a PPP \cite{baccelli}, $\Phi_1$ and $\Phi_2$ are PPPs (Note that they are \emph{not} independent) with respective densities $a\lambda $ and $(1-b)\lambda$. Moreover, we claim that $\rho=a-b$. To see this, consider the independently-marked point process $\imphi=\sum_{k\geq 0} \delta_{(X_k,\,U_k)}$, where $\delta$ is the Dirac measure, and $\set{U_k}$ are the IID marks, and $U_k\sim\mathcal{U}(0,1)$, $\forall k$. Given sets $A$ and $B$ in the Euclidean plane, and $I$, $J\subset [0,1]$, the intensity measure and the second moment measure of $\imphi$ are given in \cite{baccelli} as 
\begin{align*}
\imm(A\times I) &= \int_A\int_I \diff u \,\Lambda(\diff x) \\
				&= \immexpr \numberthis \label{ppp_mean}, \\
\imsmm\left((A\times I)\times(B\times J)\right) &=\immof{A\times I}\immof{B\times J} \\
&+\immof{(A\cap B)\times(I\cap J)} \numberthis \label{ppp_second},
\end{align*}
where $\vol{\,\cdot\,}$ is the Lebesgue measure of a set taken with respect to the appropriate number of dimensions. Note that ~\eqref{ppp_second} holds true: the ground (unmarked) point process $\Phi$ is a PPP on $\mathbb{R}^2\times[0,1]$, then the independently marked point process $\tilde{\Phi}$ is also a PPP \cite{baccelli}.
Now let $I=[0,\,a]$ and $J=[b,\,1]$. The correlation between the number of sites of $\netw$ in a set $A$ and $\nett$ in a set $B$ is
\begin{align*}
\Expect{\Phi_1(A)\Phi_2(B)} &= \displaystyle \expectno{\sum_{X_k\in\Phi\cap A}\indic{U_k\leq a}  \sum_{X_l\in\Phi\cap B}\indic{U_l>b}} \\
&= \displaystyle \expectno{\sum_{X_k\in\Phi}\indic{X_k\in A}\indic{U_k\leq a}  \sum_{X_l\in\Phi}\indic{X_l\in B}\indic{U_l>b}} \\
&= \displaystyle \expectno{\sum_{X_k\in\Phi}\sum_{X_l\in\Phi} \indic{X_k\in A}\indic{U_k\leq a} \indic{X_l\in B}\indic{U_l>b} } \numberthis \label{eqtonelli}\\
&= \displaystyle \expectno{\int\displaylimits_{A\times I\times B\times J}\imphi^{(2)}(\diff(x_1,u_1,x_2,u_2))} \\
&= \displaystyle \hphantom{\expectno{}} \int\displaylimits_{A\times I\times B\times J}\imsmmof{\diff(x_1,u_1,x_2,u_2)} \numberthis \label{eqcampbell} \\
&=  a(1-b)\lambda^2 \vol{A}\vol{B}  \\
&\phantom{=} + (a-b)\lambda\vol{A\cap B} \numberthis \label{eqfinal},
\end{align*}
where $\imphi^{(2)}=\sum_{k,\;l\geq 0}\delta_{(X_k,U_k,X_l,U_l)}$ is the \textit{second power} of $\tilde{\Phi}$ \cite{baccelli}. Step \eqref{eqtonelli} follows from Tonelli's theorem, step \eqref{eqcampbell} follows from Campbell's mean value formula for the second power of a point process, and step \eqref{eqfinal} is the result of integrating with respect to the second moment measure given in \eqref{ppp_second}. Hence, $\rho$ reduces to 
\begin{align*}
\rho &= \frac{\Expect{\Phi_1(A)\Phi_2(B)}-\expect{\Phi_1(A)}\expect{\Phi_2(B)}}{\expect{\Phi(A\cap B)}} \\
     &= \frac{(a-b)\lambda\vol{A\cap B}}{\lambda\vol{A\cap B}} \\
		 &= a-b.
\end{align*}

\bibliographystyle{IEEEtran}
\bibliography{IEEEabrv,ref}
\end{document}